\documentclass[11pt]{article}
\pdfoutput=1
\usepackage{amsmath, amssymb, graphics, graphicx}
\usepackage{textcomp, subfigure}
\usepackage{color}
\usepackage[final]{pdfpages}

\definecolor{linkcolor}{rgb}{0,0,0.6} 
\usepackage[pdftex,colorlinks=true, pdfstartview=FitV, linkcolor= linkcolor, citecolor= linkcolor, urlcolor= linkcolor, hyperindex=true,hyperfigures=true]{hyperref} 

\newcommand{\bV}{\mathbf{V}}

\newcommand{\average}[1]{\left<{#1}\right>}
\newcommand{\p}[1]{\left({#1}\right)}
\newcommand{\pq}[1]{\left[{#1}\right]}

\newcommand{\smeq}{\sigma_{m,\mathrm{eq}}}

\newcommand{\E}{\mathrm{e}}
\newcommand{\D}{\mathrm{d}}

\newcommand{\derpart}[2]{\frac{\partial #1}{\partial #2}}

\newcommand{\ii}{\imath}

\newcommand{\Ll}{\mathcal L_\lambda}
\newcommand{\Lt}{\tilde {\mathcal L}_\lambda}

\newcommand{\LlQ}{\mathcal K_\lambda}
\newcommand{\tLlQ}{\mathcal {\widetilde K}_\lambda}
\newcommand{\LlQi}{\mathcal K_{\beta_{12}-\lambda}}

\begin{document}
\title{\bf   Statistical properties of the energy exchanged between two heat baths coupled by thermal fluctuations}
%
\author{ S. Ciliberto$^1$,  A.Imparato$^2$, A. Naert$^1$, M. Tanase $^1$ \\
1 Laboratoire de Physique,  \'Ecole Normale Sup\'erieure,  C.N.R.S. UMR5672 \\ 46 All\'ee d'Italie, 69364 Lyon, France \\
2 Department of Physics and Astronomy, University of Aarhus\\ Ny Munkegade, Building 1520, DK--8000 Aarhus C, Denmark}

\maketitle

\begin{abstract}

{We study both experimentally and theoretically the statistical properties of the energy exchanged between two electrical conductors, kept at different temperature by two different heat reservoirs, and coupled by the electric thermal noise. Such a system is ruled by the same  equations as two Brownian particles  kept at different temperatures and  coupled by an elastic force.
We measure  the  heat flowing between the two reservoirs, the thermodynamic work done by one part of the system on the other, and  we show that these quantities exhibit a long time fluctuation theorem. Furthermore, we evaluate the  fluctuating entropy, which satisfies  a conservation law. These experimental  results are fully justified by the  theoretically analysis. Our results give 
more insight into the energy transfer in the famous Feymann ratchet widely studied theoretically but
never in an experiment.}
\end{abstract}
PACS:{05.40.-a, 05.70.-a, 05.70.Ln}

\newpage
\pagestyle{plain}

\noindent\hrulefill

\tableofcontents

\noindent\hrulefill

\section{Introduction}
In the study of the out-of-equilibrium  dynamics of small systems (Brownian particles\cite{bli06,Jop,Ruben,Evans02}, molecular motors
\cite{kumiko}, small devices \cite{Ciliberto}, etc.)  the role of  thermal fluctuations is central.  Indeed the thermodynamics variables, such as work, entropy  and heat, fluctuate and the study of their statistical properties is important as it can provide several constrains on the system design and mechanisms\cite{sek10,Seifert_2012}. In recent years several experiments have analyzed  systems in contact with a single heat bath and driven out of equilibrium by external forces \cite{bli06,Jop,Ruben,Evans02,kumiko,Ciliberto,alb08, alb08a,seifbech2012}. On the other hand  the important case in which the system is driven out of equilibrium by a temperature gradient and the energy
exchanges are  produced only by the thermal noise has  been analyzed in many theoretical studies on model systems
 \cite{Deridda,Jarz2004,VandenBroeck,Visco,evans_temp,Villamania,alb11,alb2} but only a few times in very recent  experimental studies  because of the intrinsic difficulties of dealing with large temperature differences in small systems \cite{nostroPRL,Pekola_1}.  
 
We report here an experimental and theoretical analysis of the energy exchanged between two conductors  kept at different temperature and coupled by the electric thermal noise. This system is probably the simplest one to test recent ideas of stochastic thermodynamics, but in spite of its simplicity the interpretation of the observations proves far from elementary. We determine experimentally  the heat flux, the out of equilibrium variance as functions of the temperature difference,  and a conservation law for the fluctuating entropy, which we justify theoretically.
We show that our system can be mapped into a mechanical one, where two Brownian particles are kept at different temperatures and  coupled by an elastic force 
\cite{VandenBroeck,Villamania,alb2}.  Thus our study  gives more insight into the properties of the heat flux, produced by mechanical coupling,  in the famous Feymann ratchet \cite{Feymann,Smoluchowski} widely studied theoretically \cite{VandenBroeck} but never in an experiment. Our results  set strong constrains on the energy exchanged between coupled nano-systems kept at different temperature. 
Therefore our investigation has implications well beyond the simple system we consider here.  \\
The system analyzed in this article  is inspired by the proof developed by  Nyquist \cite{Nyquist}, who gave, in 1928, a theoretical explanation of the measurements of Johnson \cite{Johnson} on the thermal  noise voltage  in conductors. 
Nyquist's explanation is based on equilibrium thermodynamics and considers the power exchanged by two electrically coupled  conductors, which are at  same temperature $T$ in an adiabatic environment.  Imposing the condition  of thermal equilibrium 
he concluded correctly that the thermal noise voltage across a conductor of resistance $R$ has a power spectral density  
$|\tilde \eta_\omega|^2= 4 \ k_B \, T \, R$, i.e.  the Nyquist noise formula where 
$k_B$ is the Boltzmann constant and  $T$ the temperature of the conductor. Notice that, in 1928,  many years before the proof of the fluctuation dissipation theorem (FDT), this was the second example, after the Einstein relation for Brownian motion, relating  the  dissipation of a system to the amplitude of the thermal noise. Specifically, in the Einstein relation it is the viscosity of the fluid which is related to the variance of the Brownian particles positions, whereas in the Nyquist equation it is the variance of the voltage across the conductor which is proportional to its resistance.
Surprisingly, since 1928 nobody has analyzed the consequences of keeping
the two resistances, used in the Nyquist's proof, at two different
temperatures, when the Nyquist's equilibrium condition cannot be used.  
One is thus interested in measuring the statistical properties of  the energy exchanged between  the  two conductors via the electric coupling of the two thermal noises. In this article we address this question both experimentally and theoretically and show  the analogy with  two Brownian particles kept at different temperatures and  coupled by an elastic force. The key feature in the system we consider, is that the coupling between the  two reservoirs   is obtained  only by either electrical or mechanical  thermal fluctuations. 

In a recent letter \cite{nostroPRL} we presented  several experimental results and we briefly sketched the theoretical analysis concerning the system we consider in the present paper. In this extended article we want to give a full description of the  theoretical analysis and present new experimental results and the details of the calibration procedure.

The paper is organized as follows: in section \ref{stoch:sec} we describe the experimental apparatus and the stochastic equations governing the relevant dynamic and thermodynamic quantities.  We also discuss the analogy with two coupled Brownian particles. In section \ref{fluct:sec} we develop the theoretical analysis on the fluctuations of the different forms of energy flowing across the system, and discuss the corresponding fluctuation theorems. In section \ref{exp:sec} we discuss the data analysis and the main experimental results on fluctuation theorems. Furthermore, we show experimental data confirming the validity of an entropy conservation law holding at any time. Finally we conclude in section \ref{concl:sec}.

\begin{figure}[h]
     \centering
          \includegraphics[width=0.4\textwidth]{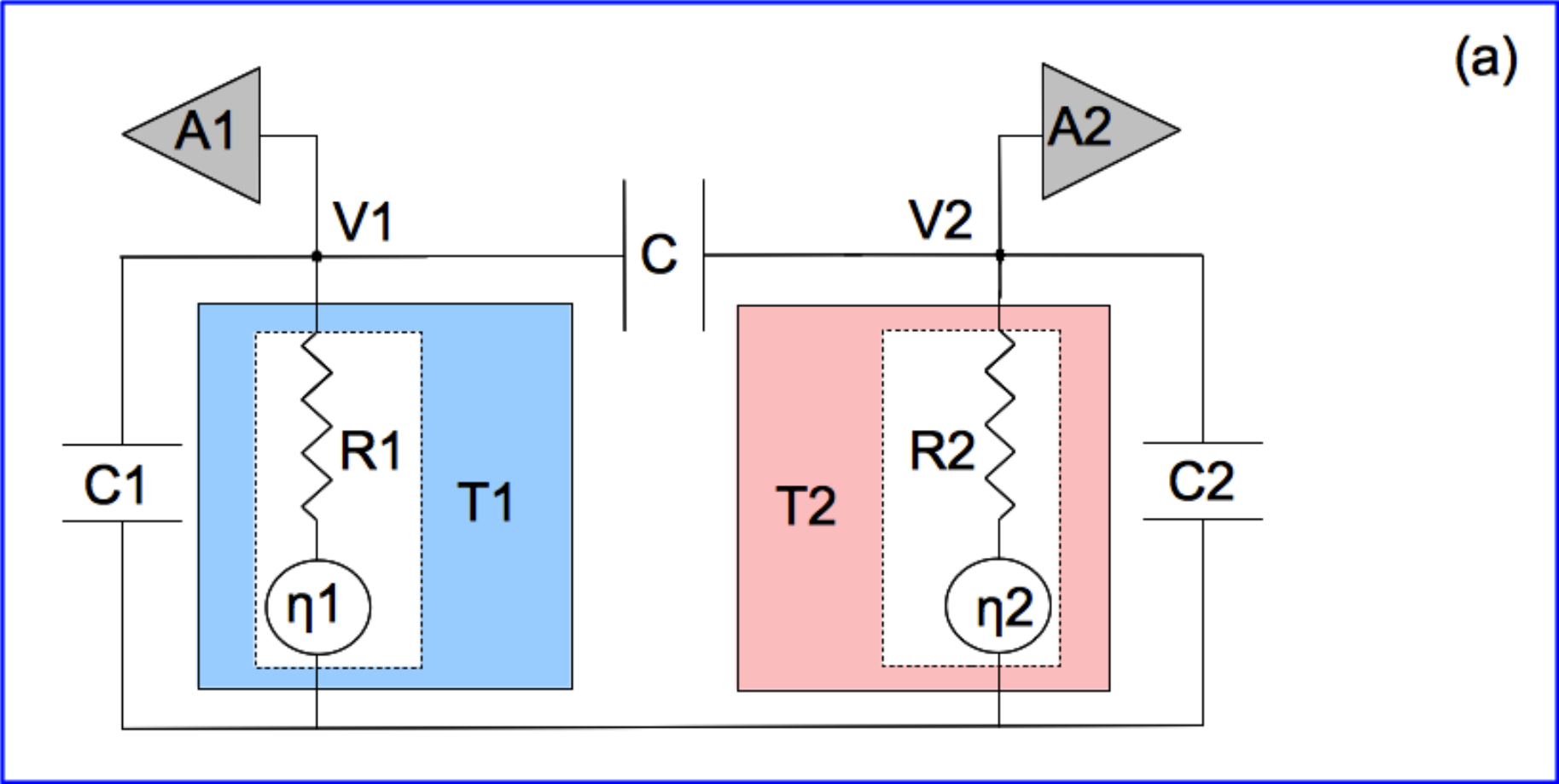}
          \includegraphics[width=0.4\textwidth]{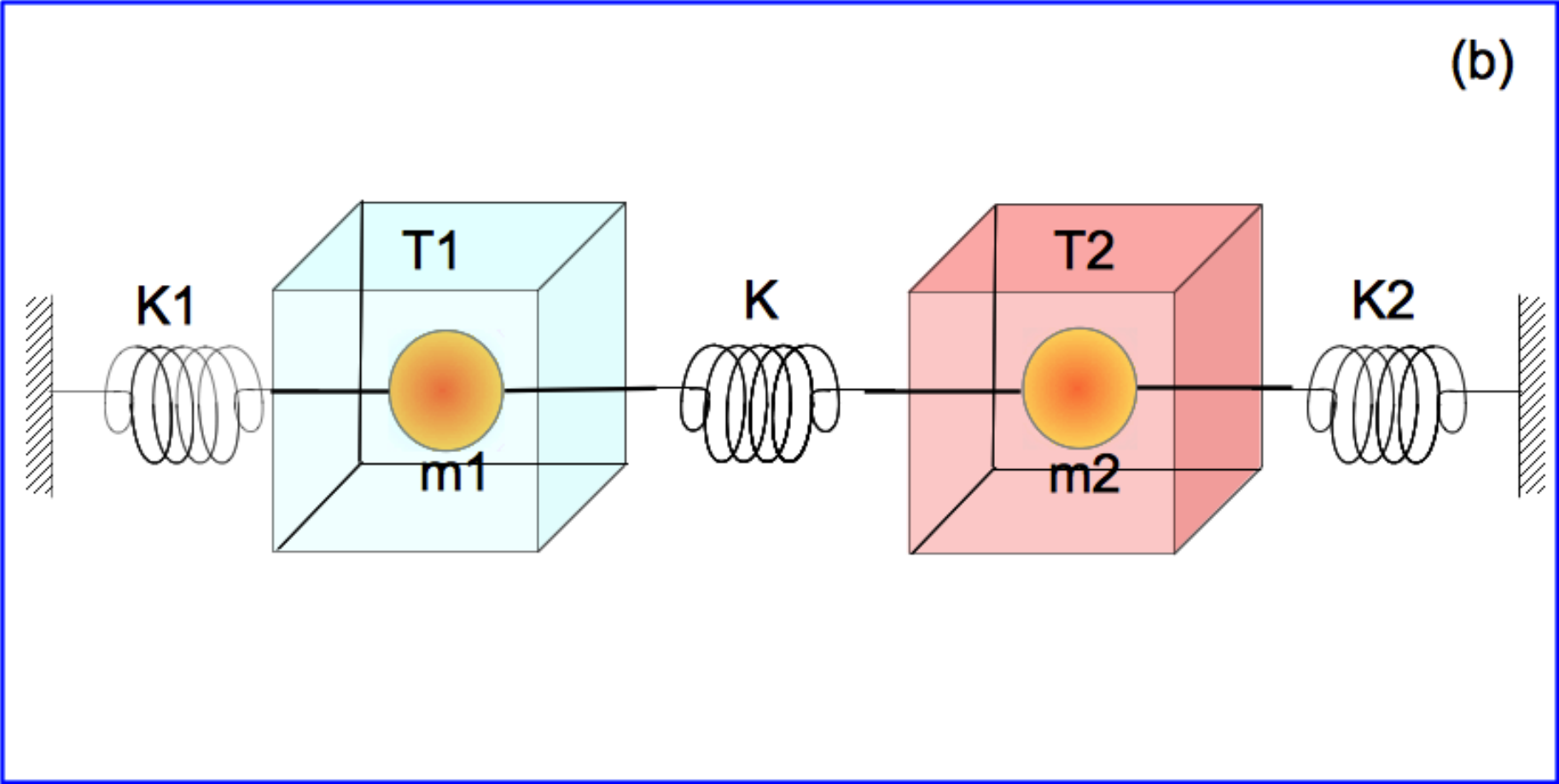}
     \caption{ a) Diagram of the circuit. The resistances $R_1$ and $R_2$ are kept at temperature $T_1$ and $T_2=296K$ respectively. They are coupled via the capacitance $C$. The capacitances $C_1$ and $C_2$ schematize the capacitance of the cables and of the amplifier inputs. 
The voltages  $V_1$ and $V_2$ are amplified  by the two low noise amplifiers $A_1$ and $A_2$ \cite{RSI}.
 b) The circuit  in  a) is  equivalent to two Brownian particles ($m_1$ and $m_2$) moving inside two different heat baths at $T_1$ and $T_2$. The two particles are trapped by two elastic potentials of stiffness $K_1$ and $K_2$ and coupled by a spring of stiffness $K$ (see text and eqs.\ref{lan1},\ref{lan2}).
}
     \label{fig:circuit}
\end{figure}

\section{Experimental set-up and stochastic variables}\label{stoch:sec}
Our experimental set-up is sketched in fig.\ref{fig:circuit}a). It is constituted by two resistances $R_1$ and $R_2$, which are kept at different temperature $T_1$ and $T_2$ respectively. These temperatures are controlled  by thermal baths and $T_2$ is kept fixed at $296K$ whereas $T_1$  can be set at a value  between $296K$ and $88K$  
using the stratified vapor above a liquid nitrogen bath.  
In the figure, the two resistances have been drawn with their associated thermal noise generators $\eta_1$ and $\eta_2$, whose power spectral densities  are given by the Nyquist formula $|\tilde \eta_m|^2= 4 k_B R_mT_m$, with $m=1,2$ (see  eqs.~(\ref{lan1})-(\ref{lan2})  ).
   The coupling capacitance $C$ controls the electrical power exchanged between the resistances  and as a consequence the energy exchanged between the two baths. {No other coupling exists between the two resistances which are inside two separated screened boxes}. 
The quantities $C_1$ and $C_2$ are the capacitances of the circuits and the cables. 
Two extremely  low noise amplifiers  $A_1$ and $A_2$ \cite{RSI} measure the voltage $V_1$ and $V_2$ across  the resistances $R_1$ and $R_2$ respectively. {All the relevant quantities considered in this paper can be derived by the measurements of $V_1$ and $V_2$, as discussed below}. 

\subsection{ Stochastic equations for the voltages}
We now proceed to derive the equations for the dynamical variables $V_1$ and $V_2$.
Furthermore, we will discuss how our system can be mapped onto a system with two interacting  Brownian particles, in the overdamped regime,  coupled  to two different temperatures,  see fig.~\ref{fig:circuit}-b).   
Let $q_m$ ($m=1,2$) be the charges that have flowed through the resistances $R_m$, so  the instantaneous current flowing through them is $i_m=\dot q_m $. 
 A  circuit analysis shows that the equations for the charges are: 
 \begin{eqnarray}
 R_1 \dot q_1&=&- q_1\, {C_2 \over X}+  (q_2-q_1){C \over X} + \eta_1  \label{lan1}\\
 R_2 \dot  q_2&=&- q_2\, {C_1  \over X}+  (q_1-q_2){C \over X} + \eta_2 \label{lan2}
\end{eqnarray}
where   $\eta_m$ is the usual white noise: $\average{\eta_i(t)\eta_j(t')}=2 \delta_{ij} {k_BT_i R_j} \delta(t-t')$, and where we have introduced the quantity  $X=C_2\, C_1+C\, (C_1 \, +C_2) $. 
Eqs.~\ref{lan1} and \ref{lan2} are the same of  those for the two coupled Brownian particles sketched in fig.\ref{fig:circuit}b)  when one regards 
$q_m$ as the displacement of the particle $m$, $i_m$ as its velocity,  $K_m=1/C_m$ as the stiffness of the spring $m$, $K=1/C$ as the coupling spring and $R_m$ the viscosity.  The analogy with the Feymann ratchet can be made by assuming as done in ref.\cite{VandenBroeck} that the particle $m_1$ has an asymmetric shape and on average moves faster in one direction than in the other one.


We now rearrange eqs.~(\ref{lan1})-(\ref{lan2}) to obtain the Langevin equations for the voltages, which will be useful in the following discussion.
The relationships between the measured voltages and the charges are: 
\begin{eqnarray}
q_1&=& (V_1-V_2) \, C + V_1\, C_1  \label{eq_q1}\\
q_2&=& (V_1-V_2) \, C - V_2\, C_2  \label{eq_q2}.
\end{eqnarray}
By plugging eqs.~(\ref{eq_q1})-(\ref{eq_q2}) into eqs.(\ref{lan1})-(\ref{lan2}), and rearranging terms, we obtain
\begin{eqnarray}
 (C_1+C) \dot V_1&=& C \dot V_2 + \frac{1}{R_1}(\eta_1-V_1)\label{lan01},\\
(C_2+C) \dot V_2&=& C \dot V_1 +\frac{1}{R_2}(\eta_2-V_2)\label{lan02}.
\end{eqnarray}

We rearrange these equations in a standard form, and obtain
\begin{eqnarray}
\dot V_1&=&f_1(V_1,V_2) + \sigma_{11}\eta_1+\sigma_{12} \eta_2=f_1(V_1,V_2)+\xi_1 \label{lanV1}\\
\dot V_2&=&f_2(V_1,V_2) + \sigma_{21}\eta_1+\sigma_{22} \eta_2=f_2(V_1,V_2)+\xi_2 \label{lanV2}
\end{eqnarray}
where the ``forces'' acting on the circuits read 
\begin{eqnarray}
f_1(V_1,V_2)&=& -\pq{\frac{(C+C_2)  V_1}{R_1 X}+\frac{C V_2}{R_2 X} }\label{eqf1},\\
f_2(V_1,V_2)&=&-\pq{\frac{C  V_1}{R_1 X}+\frac{(C+C_1) V_2}{R_2 X} }\label{eqf2},
\end{eqnarray}
the coefficients $\sigma_{ij}$ read
\begin{eqnarray}
\sigma_{11}&=&\frac{C_2 +C}{X R1} \nonumber\\
R_2 \sigma_{12}&=&R_1 \sigma_{21}=\frac C X \nonumber\\
\sigma_{22}&=&\frac{C_1 +C}{X R_2},\nonumber
\end{eqnarray} 
and the noises $ \xi_i$ introduced in eqs.~(\ref{lanV1})-(\ref{lanV2}) are now correlated  $\average{\xi_i\xi'_j}=2 \theta_{ij} \delta(t-t')$, where 
\begin{eqnarray}
\theta_{11} &=& \frac{k_B T_1 (C_2 + C)^2}{R_1X^2} + 
   \frac{k_B T_2 C^2}{R_2 X^2} \label{theta1},\\
\theta_{12} &=&  \frac{k_B T_1 C (C_2 + C)}{R_1 X^2} + \frac{k_B T_2 C (C_1 + C)}{R_2 X^2} \label{theta12},\\
\theta_{22} &=&  \frac{k_B T_1 C^2}{R_1 X^2} + \frac{k_B T_2 (C_1 + C)^2}{R_2 X^2}\label{theta2},
\end{eqnarray} 
and $\theta_{12}=\theta_{21}$.
\subsection{ Stochastic equations for work and heat exchanged between the two circuits}
Two important quantities can be identified in the circuit depicted in fig.~\ref{fig:circuit}: the electric power dissipated in each resistor, and the work exerted by one circuit on the other one.
We start by considering the first quantity $Q_m$, defined through the dissipation rate $\dot Q_m=V_m i_m$, where $i_m$ is the current flowing in the resistance $m$.
As the voltages $V_m$ can be measured, one can obtain the currents as $i_m= i_C-i_{C_m}$, where 
\begin{equation}
i_C=C \p{\dot V_2-\dot V_1}, \qquad i_{C_m}=C_m \dot V_m,
\end{equation} 
are the current flowing in the capacitance $C$ and in $C_m$, respectively.
Thus the total energy dissipated by the resistance $m$ in a time interval $\tau$ reads
\begin{equation}
Q_{m,\tau}=\int_{t_0}^{t_0+\tau} i_m(t) V_m(t)\D t=\int_{t_0}^{t_0+\tau} V_m\pq{C \dot V_{m'}-(C_m+C)\dot V_m}\D t, 
\label{eq:dQ}
\end{equation} 

We see that in equation (\ref{eq:dQ}) we can isolate the term $C V_{m}\dot V_{m'}\D t$, denoting the work rate done by one circuit on the other one, 
from which we obtain the integrated quantities  
\begin{equation}
W_{m,\tau}=\int_{t_0}^{t_0+\tau} C V_{m}(t) \dot V_{m'}(t)\D t.
\label{eq:W}
\end{equation} 
and 
\begin{equation}
\Delta U_{m,\tau}= {1\over 2} (C_m+C) (V_m^2(t+\tau)-V_m^2(t)
\label{eq:DeltaUm}
\end{equation}
The quantities  $W_{m,\tau}$ can  be thus  identified as the thermodynamic work performed by the circuit $m'$  on $m$ \cite{Sekimoto,VanZonCil,Garnier}. 
As the two variables $V_m$ are fluctuating voltages, the derived quantities $Q_{m,\tau}$ and $W_{m,\tau}$ fluctuate too.

By plugging eqs.~(\ref{lanV1})-(\ref{lanV2}) into the definitions of dissipated energy and work, eqs.~(\ref{eq:dQ}) and (\ref{eq:W}), respectively, we obtain 
the Langevin equations governing the time evolution of the two thermodynamic quantities:
\begin{eqnarray}
\dot W_m&=&C V_m \dot V_{m'}=C V_m(f_{m'}+ \xi_{m'}), \label{wm}\\ 
\dot Q_m&=&V_m i_m=V_m\pq{C \dot V_{m'}-(C_m+C)\dot V_m}=\frac{V_m}{R_m}(V_m-\eta_m).\label{qm}
\end{eqnarray} 

It is instructive to reconsider the quantity $Q_{m,\tau}$ in terms of the stochastic energetics  \cite{sek10}.
If we introduce the circuit total potential energy, defined as
\begin{equation}
U=\frac{C_1}{2} V_1^2 + \frac C 2 (V_1-V_2)^2 +\frac{C_2}{2} V_2^2=\frac{C_2 q_1^2+ C (q1-q2)^2+C_1 q_2^2}{2 X},
\label{defU}
\end{equation} 
by noticing that eqs.(\ref{lan1})-(\ref{lan2}) can be written as $R_m \dot q_m=-\partial_{q_m} U +\eta_m$, and 
following Sekimoto \cite{sek10} we see that we can write 
the dissipated energy as 
\begin{equation}
Q_{m,\tau}=-\int_{t_0}^{t_0+\tau}\derpart{U}{q_m} \D q_m=\int_{t_0}^{t_0+\tau} \frac{V_m}{R_m}(V_m-\eta_m) \D t,
\end{equation} 
where we have expressed the charges in terms of the voltages by inverting eqs.~(\ref{eq_q1})-(\ref{eq_q2}).
With the analogy of the Brownian particles, depicted in fig.~\ref{fig:circuit}-b),  we see that our definition of dissipated energy $Q_m$ corresponds exactly to the work performed by the viscous forces and by the bath on
the particle $m$, and it is consistent with the stochastic thermodynamics
definition  \cite{sek10,Seifert_2012,alb2,Sekimoto,VanZonCil,Garnier,alb1}. 
Thus, the quantity $Q_{1,\tau}$ ($Q_{2,\tau}$) can  be interpreted as the heat flowing from the reservoir 2 to the reservoir 1 (from 1 to 2), in the time interval $\tau$, as an effect of the temperature difference. 

Hence we have derived the set of Langevin equations, describing the time evolution of the dynamical variables for $V_m$,  and of the thermodynamic variables $Q_m$ and $W_m$.
One expects that both these thermodynamic quantities satisfy a fluctuation theorem (FT)  of the type \cite{Jarz2004,Visco,alb2,alb1,EvansFT,gallavotti} 
\begin{equation}
\ln {P(E_{m,\tau})\over P(-E_{m,\tau})} =  \beta_{12} \ E_{m,\tau} \Sigma(\tau)
\label{eq_Pq}
\end{equation}
where $E_{m,\tau}$ stands either for $W_{m,\tau}$ or $Q_{m,\tau}$, $\beta_{12} =(1/T_1-1/T_2)/k_B$ and $\Sigma(\tau)\rightarrow 1$ for $\tau \rightarrow\infty$. 
In order to prove this relation, we need to discuss the statistics of the fluctuations of the quantity of interests, namely $V_m$, $W_m$, and $Q_m$.
\section{Fluctuations of $V_m$, $W_m$ and $Q_m$}\label{fluct:sec}
\subsection{Probability distribution function for the voltages}
We now study the joint probability distribution function (PDF) $P(V_1,V_2,t)$, that the system at time $t$ has a voltage drop $V_1$ across the resistor $R_1$ and a  voltage drop $V_2$ across the resistor $R_2$.
As the time evolution of $V_1$ and $V_2$ is described by the Langevin equations (\ref{lanV1})-(\ref{lanV2}), it can be proved that the time evolution of $P(V_1,V_2,t)$ is governed by the Fokker-Planck equation \cite{Zwa}
\begin{eqnarray}
\partial_t P(V_1,V_2,t)&= &L_0 P(V_1,V_2,t)= -\frac{\partial}{\partial V_1} \p{f_1  P }-\frac{\partial}{\partial V_2} \p{f_2  P }+ 2 \theta_{12}\frac{\partial^2}{\partial V_1 \partial V_2} P\nonumber \\
&&\qquad\qquad\qquad +\theta_{11}\frac{\partial^2}{\partial V_1^2} P +\theta_{22}\frac{\partial^2}{\partial V_2^2} P
\label{L0}
\end{eqnarray}
We are interested in the  long time steady state solution of eq.~(\ref{L0}), which is time independent $P(V_1,V_2,t\rightarrow\infty)=P_{ss}(V_1,V_2)$.
As the deterministic forces in eqs.~(\ref{lanV1})-(\ref{lanV2}) are linear in the variables $V_1$ and $V_2$, such a  steady state solution 
reads
\begin{equation}
P_{ss}(V_1,V_2)=\frac{\pi\E^{-m_{ij}V_iV_j}}{\sqrt{\det \mathbf{m}}},
\label{phi:ss}
\end{equation} 
where the sum over repeated indices is understood, and 
where the $\mathbf{m}$ matrix entries read
\begin{eqnarray}
m_{11}&=&\frac{Y \left[T_2 (C+C_1)Y+C^2 R_2 (T_1-T_2)\right]}{2k_B\pq{Y^2 T_1 T_2+C^2 R_1 R_2 (T_1-T_2)^2} },\nonumber\\
m_{12}&=&m_{21}=-\frac{Y C [(C_2 +C) R_2 T_1 + (C_1+C) R_1 T_2]}{2 k_B\pq{Y^2 T_1 T_2+C^2 R_1 R_2 (T_1-T_2)^2} }\nonumber,\\
m_{22}&=&\frac{Y \left[T_1 (C+C_2)Y-C^2 R_1 (T_1-T_2)\right]}{2k_B\pq{Y^2 T_1 T_2+C^2 R_1 R_2 (T_1-T_2)^2} },\nonumber
\end{eqnarray} 
where we have introduced the quantity
$Y=\pq{(C_1+C) R_1 +(C_2+C) R_2 }$.

Such a solution can be obtained by replacing eq.~(\ref{phi:ss}) into eq.~(\ref{L0}), and by imposing the steady state condition $\partial_t P=0$.
We are furthermore interested in the unconstrained steady state probabilities $P_{1,ss}(V_1)$, and $P_{2,ss}(V_2)$, which are obtained as follows
\begin{eqnarray}
P_{1,ss}(V_1)&=&\int \D V_2 P_{ss}(V_1,V_2)= \frac{\E^{-\frac{V_1^2}{2\sigma_{1}^2}}}{\sqrt{2 \pi \sigma_{1}^2}}\\
P_{2,ss}(V_2)&=&\int \D V_1 P_{ss}(V_1,V_2)= \frac{\E^{-\frac{V_2^2}{2\sigma_{2}^2}}} {\sqrt{2 \pi \sigma_{2}^2}}
\end{eqnarray} 
where the variances read
\begin{eqnarray}
\sigma_{1}^2&=&k_B \frac{T_1 (C+C_2) Y  +(T_2-T_1)C^2 R_1}{X Y}\label{eq:s1}\\
\sigma_{2}^2&=&k_B \frac{T_2 (C+C_1) Y  -(T_2-T_1)C^2 R_2}{X Y}\label{eq:s2}
\end{eqnarray}

\subsection{Average value and long time FT for $W_1$}\label{secFTW1}

In eqs.~(\ref{eq:dQ})-(\ref{eq:W}) $t_0$ denote the instant when one begins to measure the thermodynamic quantities. In the following we will  assume that the system is already in a steady state at that time and take $t_0=0$ for simplicity.
%
We will discuss the case of $W_1$ without loss of generality, the mathematical treatment for $W_2$ being identical.
We first notice that the dynamics of $W_1$ is described by the Langevin equation (\ref{wm}): the noise affecting  $W_1$ is $C V_1 \xi_2$, which is thus correlated to the noises $\xi_1, \, \xi_2$ affecting $V_1$ and $V_2$ through the diffusion matrix defined in eqs. (\ref{theta1})-(\ref{theta2}).
We introduce the joint probability distribution $\phi(V_1, V_2, W_1, t)$:
the time evolution of such a PDF is described by the  Fokker-Planck equation
\begin{eqnarray}
\partial_t \phi(V_1, V_2, W_1, t)&=&-\frac{\partial}{\partial V_1} \p{f_1 \phi}-\frac{\partial}{\partial V_2} \p{f_2 \phi}+\theta_{11}\frac{\partial^2}{\partial V_1^2} \phi+\theta_{22}\frac{\partial^2}{\partial V_2^2} \phi\nonumber \\
&&+2  \theta_{12}\frac{\partial^2}{\partial V_1 \partial V_2} \phi
- C \frac{\partial}{\partial W_1} \p{V_1 f_2 \phi} \nonumber \\
&&+\theta_{12} C \pq{\frac{\partial}{\partial V_1}\p{V_1 \frac{\partial}{\partial W_1} \phi}+\frac{\partial}{\partial W_1}\p{V_1 \frac{\partial}{\partial V_1} \phi}}\nonumber \\
& &+2 \theta_{22} C \frac{\partial}{\partial V_2 }\p{V_1 \frac{\partial}{\partial W_1} \phi}+\theta_{22} (C V_1)^2 \frac{\partial^2}{\partial W_1^2} \phi.
\end{eqnarray} 
We now introduce the generating function defined as $\psi(V_1, V_2, \lambda,t)=\int \D W_1 \exp(\lambda W_1) \phi(V_1, V_2, W_1,t)$, whose dynamic is described by the Fokker-Planck equation
\begin{equation}
\partial_t \psi(V_1, V_2, \lambda,t)=\mathcal L_\lambda \psi, 
\label{dtpsi}
\end{equation} 
where the operator $\mathcal L_\lambda$ reads 
\begin{eqnarray}
\mathcal L_\lambda \psi  &=&-\frac{\partial}{\partial V_1} \p{f_1  \psi }-\frac{\partial}{\partial V_2} \p{f_2  \psi }+\theta_{11}\frac{\partial^2}{\partial V_1^2} \psi +\theta_{22}\frac{\partial^2}{\partial V_2^2} \psi + 2 \theta_{12}\frac{\partial^2}{\partial V_1 \partial V_2} \psi \nonumber\\
& & - \lambda \theta_{12} C \pq{\frac{\partial}{\partial V_1}\p{V_1 \psi }+\p{V_1 \frac{\partial}{\partial V_1} \psi }}\nonumber \\
& &-2 \theta_{22} \lambda C \frac{\partial}{\partial V_2 }\p{V_1 \psi }+\lambda C V_1 (\theta_{22} \lambda C V_1 +f_2)\psi. 
\label{L_lam}
\end{eqnarray} 
For the average value of the work, after a straightforward calculation, one finds 
\begin{eqnarray}
\partial_t \average{W_1}&=&\pq{\partial_\lambda \partial_t \int \D V_1 \D V_2 \psi(V_1,V_2,\lambda,t)}_{\lambda=0}=\frac{C^2 k_B(T_2-T_1)}{XY},
\label{eqW1}
\end{eqnarray}

As we are interested in the large time limit of the unconstrained generating function, we notice that 
$\int \D V_1 \D V_2 \psi(V_1,V_2,\lambda,t)\propto \exp\pq{t \mu_0(\lambda)}$, where $\mu_0(\lambda)$ is the largest eigenvalue of the operator $\mathcal L_\lambda$. Thus, proving that the unconstrained PDF $P(W_1,\tau)=\int \D V_1 \D V_2 \phi(V_1,V_2,W_1,t)$ satisfies the FT (\ref{eq_Pq}) is equivalent to prove that $\mu_0(\lambda)$ exhibits the following symmetry:
\begin{equation}
\mu_0(\lambda)=\mu_0(-\lambda-\beta_{12}).
\label{sym_mu}
\end{equation} 
In order to prove such an equality, following \cite{alb2} we introduce the  
operator
\begin{equation}
\Lt=\E^{H} \Ll \E^{- H},
\end{equation} 
where $H(V_1,V_2)$ is some dimensionless Hamiltonian to be determined: thus 
this transformation corresponds to a ``rotation'' of the operator $\Ll$, or more precisely $\Lt$ and $\Ll$ are related by a unitary transformation.

Let's consider an eigenvector $\psi_n(V_1,V_2,\lambda)$ of the original operator $\mathcal L_\lambda$, with eigenvalue $\mu_n(\lambda)$, then 
one easily finds that the following equality holds
\begin{equation}
\Lt \E^{H} \psi_n(V_1,V_2,\lambda)=\E^{H} \Ll \E^{ -H}  \E^{ H} \psi(V_1,V_2,\lambda)=\mu_n(\lambda) \E^{ H} \psi(V_1, V_2, \lambda),\label{lrot}
\end{equation} 
thus, $\mathcal L_\lambda$ and $\Lt$ have the same eigenvalues, only the eigenvectors are ``rotated'' by the operator $\exp(H)$.
Note that eq.~(\ref{lrot}) holds for any choice of $H$.

Our goal is still to prove eq.~(\ref{sym_mu}). 
By choosing 
\begin{equation}
H=\frac{C_1  + C}{2 k_B T_1} V_1^2 -  \frac{C}{k_B T_2} V_1 V_2  + \frac{C_2+C }{2 k_B T_2}  V_2^2\label{H:def}.
\end{equation} 
one finds that 
the following equality holds
\begin{equation}
\Lt=\mathcal{L}^*_{-\lambda-\beta_{12}},
\label{ltls}
\end{equation} 
where $\mathcal{L}^*_\lambda$ is the adjoint operator of 
$\mathcal{L}_{\lambda}$.
From the above discussion we know that 
 $\mathcal L_\lambda$ and 
$\Lt$ have the same eigenvalues, while eq.~(\ref{ltls}) shows that $\Lt$ and $\mathcal{L}^*_{-\lambda-\beta_{12}}$ are the same operator, and so that $\mathcal{L}_{\lambda}$ and  $\mathcal{L}^*_{-\lambda-\beta_{12}}$ have the same spectra of eigenvalues, and in particular identical maximal eigenvalues. Thus we conclude that $\mu_0(\lambda)=\mu_0(-\lambda-\beta_{12})$, which is the FT (\ref{eq_Pq}) in the form of eq.~(\ref{sym_mu}).

\subsection{Average value and long time FT for $Q_m$}\label{secFTQ1}
We now consider the dissipated heat, defined through its time derivative, as given by eq.~(\ref{qm}).
Similarly to what we have done for $W_1$, 
we now introduce the joint PDF $\pi(V_1,V_2,Q_1,t)$, and the corresponding generating function  $\chi(V_1,V_2,\lambda,t)=\int \D Q_1 \exp(\lambda Q_1) \pi(V_1, V_2, Q_1,t)$, obtaining the Fokker-Planck equation
\begin{equation}
\partial_t \chi(V_1, V_2, \lambda,t)=\LlQ \chi, 
\end{equation} 
where the operator $\LlQ$ reads 
\begin{eqnarray}
\LlQ \chi  &=&-\frac{\partial}{\partial V_1} \p{f_1  \chi }-\frac{\partial}{\partial V_2} \p{f_2  \chi }+\theta_{11}\frac{\partial^2}{\partial V_1^2} \chi +\theta_{22}\frac{\partial^2}{\partial V_2^2} \chi + 2 \theta_{12}\frac{\partial^2}{\partial V_1 \partial V_2} \chi \nonumber\\
& & + \lambda r_{11}  \pq{\frac{\partial}{\partial V_1}\p{V_1 \chi }+\p{V_1 \frac{\partial}{\partial V_1} \chi }}\nonumber \\
& &+2 \lambda  r_{12} \frac{\partial}{\partial V_2 }\p{V_1 \chi }+\lambda V_1^2 \p{\lambda r_{22} +\frac 1{R_1}}\chi,
\label{L_lam_Q1}
\end{eqnarray} 
with 
\begin{eqnarray}
r_{11}&=& k_1 \theta_{11}+ k_2 \theta_{12},\nonumber\\
r_{12}&=& k_1 \theta_{12}+ k_2 \theta_{22},\nonumber\\
r_{22}&=& k_1^2 \theta_{11}+ k_2^2 \theta_{22}+2 k_1 k_2 \theta_{12},\nonumber\\
\end{eqnarray} 
and $k_1=(C_1+C)$, $k_2=-C$.
Thus, after a straightforward calculation, we obtain the heat rate as given by 
\begin{eqnarray}
\partial_t \average{Q_1}&=&\pq{\partial_\lambda \partial_t \int \D V_1 \D V_2 \chi(V_1,V_2,\lambda,t)}_{\lambda=0}=\frac{C^2 k_B (T_2-T_1)}{XY}.
\label{dtQ1}
\end{eqnarray} 
The last result is identical to eq.~(\ref{eqW1}), thus the  averages
of the two energies are equal $\average{W_1(t)}=\average{Q_1(t)}$.
This can be easily understood by noticing that $Q_{m,\tau}$ and $W_{m,\tau}$ differ by a term proportional to $\int V_m \dot V_m \D t'=\Delta V_m^2$, which vanishes on average in the steady state.

We can now relate the variance of $V_1$ and $V_2$ to the mean heat flux:
using eq.(\ref{dtQ1}) we can express eq.~(\ref{eq:s1}) and eq.~(\ref{eq:s2})  in the following way:   
\begin{eqnarray}
\sigma_{m}^2&=&\smeq^2 +<\dot Q_m> R_m\label{eq:s1_Q}
\end{eqnarray} 
where $\smeq^2=k_BT_m (C+C_{m'})/ X $ is the equilibrium value of $\sigma_{m}^2$, when $T_m=T_{m'}$, and so  $<\dot Q_m>=0$. 
Equation (\ref{eq:s1_Q}) represents  an extension to the two temperatures case of the Harada-Sasa relation \cite{harada}, which relates 
the difference of the equilibrium and out-of-equilibrium power spectra to the heat fluxes.

Following the same route described in section \ref{secFTW1}, we now want to prove the FT for the unconstrained heat distribution   PDF $P(Q_1,\tau)=\int \D V_1 \D V_2 \pi(V_1,V_2,Q_1,t)$ satisfies the FT (\ref{eq_Pq}),
which is equivalent to the requirement
\begin{equation}
\nu_0(\lambda)=\nu_0(\beta_{12}-\lambda),
\label{eqnu0}
\end{equation} 
where $\nu_0(\lambda)$ is the largest eigenvalue of the operator $\LlQ$, and so in the large time limit one expects $\int  \D V_1 \D V_2 \chi(V_1,V_2,\lambda,t)\propto \exp\pq{\nu_0(\lambda) t}$ .
We introduce the transformation

\begin{equation}
\tLlQ=\E^{H} \LlQ \E^{- H},
\end{equation} 
where the ``Hamiltonian'' generator of the transformation reads
$H=U/(k_B T_2) $
and where $U$ is given by eq.~(\ref{defU}).
We then find, after a lengthy but straightforward calculation that $\tLlQ=\LlQi^*$
where $\LlQ^*$ is the adjoint operator of 
$\LlQ$. Thus we infer that  $\LlQ$ and  $\LlQi^*$ have the same spectra of eigenvalues, and in particular identical maximal eigenvalues, and so eq.~(\ref{eqnu0}) and the FT (\ref{eq_Pq}) follow.

\section{Analysis of the experimental data  }\label{exp:sec}

\subsection{Experimental details} 
The electric systems and amplifiers are inside a Faraday cage and mounted on a floating optical table to reduce mechanical  and acoustical noise. The resistance $R_1$, which  is cooled by liquid Nitrogen vapors, changes of less than $0.1 \%$ in the whole temperature range. Its temperature is measured by a PT1000 which is inside the same shield  of $R_1$. The signal $V_1$ and $V_2$ are amplified by two custom designed JFET amplifiers \cite{RSI} with an input current of $1pA$ and a  noise of $0.7nV/\sqrt{Hz}$ at frequencies larger than $1Hz$ and increases at $8nV/\sqrt{Hz}$ at $0.1Hz$, see fig.~\ref{fig:spectre}.  The resistances $R1$ and $R2$ have been used as input resistances of the amplifiers. The two signals  $V_1$ and $V_2$  are amplified   $10^4$  times and the amplifier outputs are filtered (at $4kHz$ to avoid aliasing) and acquired at $8kHz$ by  24 bits-ADC.   We used different sets of $C_1,C_2$ and $C$. The values of  $C1$ and $C2$ are essentially set by the input capacitance of the amplifiers and by the cable length $680pF<C1 <780pF$  and $400pF<C_2<500pF$. Instead $C$ has been changed from $100pF$ to $1000pF$. 
In the following we will take  $C=100pF,  C_1=680pF, C_2=420pF$ and $R_1=R_2=10M\Omega$, if not differently stated. The longest characteristic time of the system is $Y=\pq{(C_1+C) R_1 +(C_2+C) R_2 }$ which for the mentioned values of the parameters is : $Y=13\,$ms.

\subsubsection{Check of the calibration}
When $T_1=T_2=296K$ the system is in equilibrium and exhibits no net energy flux  between the two reservoirs. 
This is indeed the condition imposed  by Nyquist to prove his formula, and we use it  {to check all the values of the circuit parameters.} 
{Applying the Fluctuation-Dissipation-Theorem (FDT) to the circuit in fig.\ref{fig:circuit}a), one finds the  Nyquist's expression for the variance  of $V_1$ and $V_2$ at equilibrium, which reads  $ \smeq^2(T_m)= {k_B T_m (C+C_{m'})/X }$ with $X=C_2\, C_1+C\, (C_1 \, +C_2) $,  $m'=2$ if $m=1$ and $m'=1$ if $m=2$.}  For example one can check that at $T_1=T_2=296$ K, using  the above mentioned values of the capacitances and resistances,  the predicted equilibrium standard deviations of  $V_1$ and $V_2$ are  $2.33 \mu V$ and $8.16 \mu V$ respectively. These are indeed the measured values  with an accuracy better than $ 1 \% $.
The equilibrium spectra of $V_1$ and $V_2$  at $T_1=T_2$ used for calibration of the capacitances  are: 
\begin{eqnarray} 
 Sp_1(\omega) &=& {4 k_BT_1 \, R_1  [1+\omega^2(C^2 R_1 R_2+R_2^2 (C_2+C)^2)]  \over   
(1-\omega^2 \, X \, R_1R_2)^2+\omega^2 Y^2} \label{eq:equilibrium_spectra_1}, \\
Sp_2(\omega)&=&{4 k_B T_2 \, R_2  [1+\omega^2(C^2 R_1 R_2+R_1^2 (C_1+C)^2) ]\over   
(1-\omega^2 \, X \, R_1R_2)^2+\omega^2 Y^2}.
\label{eq:equilibrium_spectra_2}
\end{eqnarray}
This spectra can be easily obtained by applying FDT  to the circuit of fig.\ref{fig:circuit}.  

The two computed spectra are compared to the measured ones  in fig.~\ref{fig:spectre}a).   This comparison allows us to check the values of the capacitances $C_1$ and $C_2$ which depend on the cable length. We see that the agreement between the prediction and the measured power spectra  is excellent and the global error on calibration is of the order of $1\%$. This corresponds exactly to the case discussed by Nyquist in which the two resistances at the same temperature are exchanging energy via an electric circuit ($C$ in our case).

\subsubsection{Noise spectrum of the amplifiers}
\begin{figure}[h]
     \centering
     \includegraphics[width=0.3\textwidth]{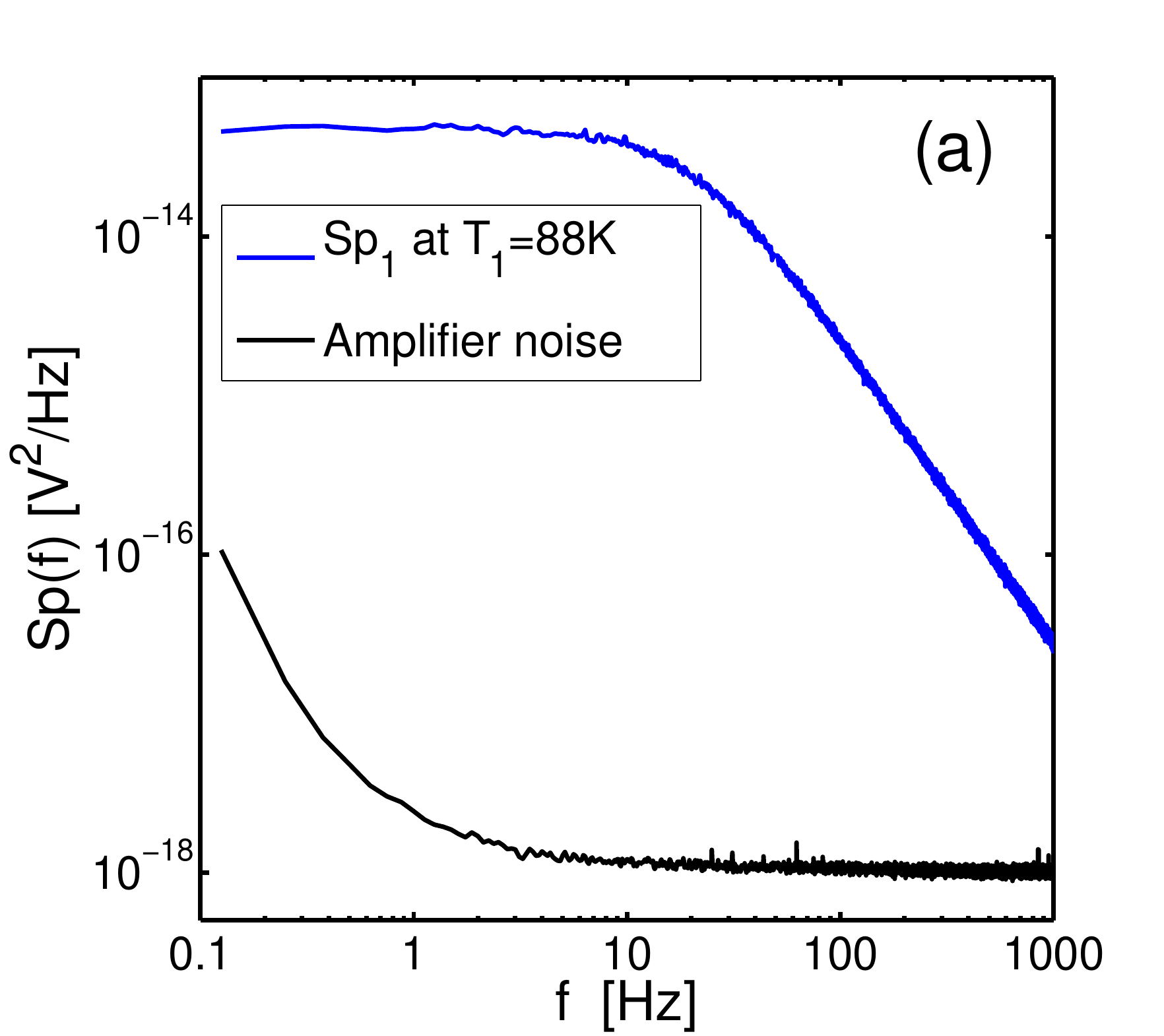}
          \includegraphics[width=0.3\textwidth]{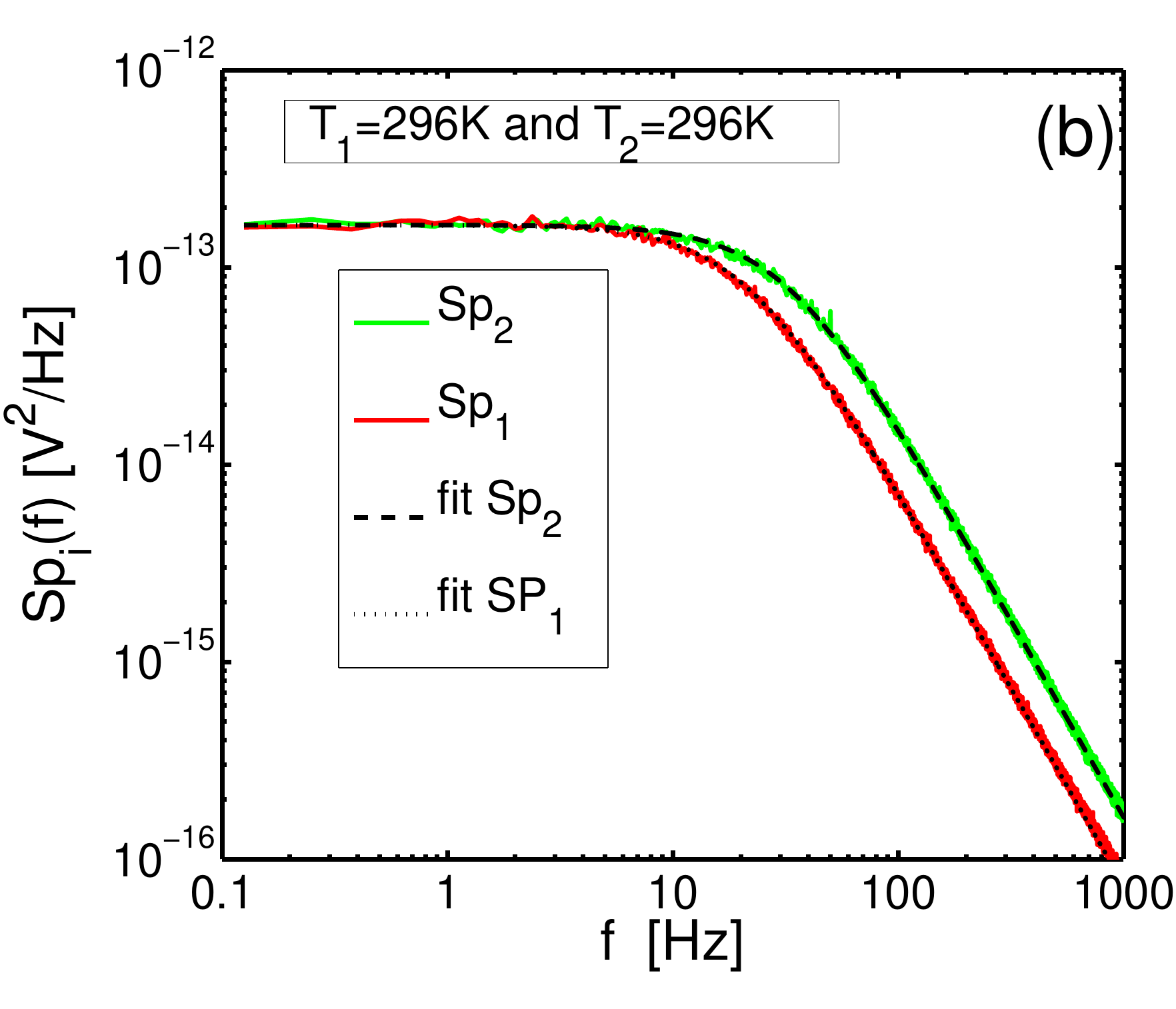}
  \caption{ a) The power spectra $Sp_1$ of  $V_1$  measured  at $T_1=88K$  (blue line) ($C=100pF,C_1=680pF,C_2=430pF$) is compared to the spectrum of the amplifier noise. b) The equilibrium spectra $Sp_1$(red line) and $Sp_2$ (green line) measured at 
    $ T_1=T_2=296K$ are compared with prediction of eqs.~(\ref{eq:equilibrium_spectra_1}) and (\ref{eq:equilibrium_spectra_2}) in order to check the values of the capacitances ($C_1,C_2$).    
     }
     \label{fig:spectre}
     \end{figure}

The noise spectrum of the amplifiers $A_1$ and $A_2$  (Fig.\ref{fig:circuit}a), measured with a short circuit at the inputs, is plotted in fig.\ref{fig:spectre}a) and compared with the spectrum $Sp_1$ of $V_1$ at $T_1=88K$.
We see that the useful signal is several order of magnitude larger than the amplifiers noise. 

\subsection{The statistical properties of  $V_m$ }

\subsubsection{The power  spectra and the variances of $V_m$  out-of-equilibrium}

 When $T_1\ne T_2$  the power spectra of $V_1$ and $V_2$ are: 
\begin{eqnarray} 
Sp_1(\omega) &=& {4 k_BT_1 \, R_1  [1+\omega^2(C^2 R_1 R_2+R_2^2 (C_2+C)^2)]  \over   
(1-\omega^2 \, X \, R_1R_2)^2+\omega^2 Y^2} + {4 k_B(T_2-T_1) \,  \omega^2 \, C^2 R_1^2 R_2  \over   
(1-\omega^2 \, X \, R_1R_2)^2+\omega^2 Y^2}   \label{eq.spectra_out1}\\
Sp_2(\omega) &=& {4 k_BT_2 \, R_2  [1+\omega^2(C^2 R_1 R_2+R_1^2 (C_1+C)^2) ]\over   
(1-\omega^2 \, X \, R_1R_2)^2+\omega^2 Y^2}+ {4 k_B(T_1-T_2) \,  \omega^2 \, C^2 R_2^2 R_1 \over   
(1-\omega^2 \, X \, R_1R_2)^2+\omega^2 Y^2}
 \label{eq.spectra_out2}
\end{eqnarray}

\begin{figure}[h]
     \centering
           \includegraphics[width=0.3\textwidth]{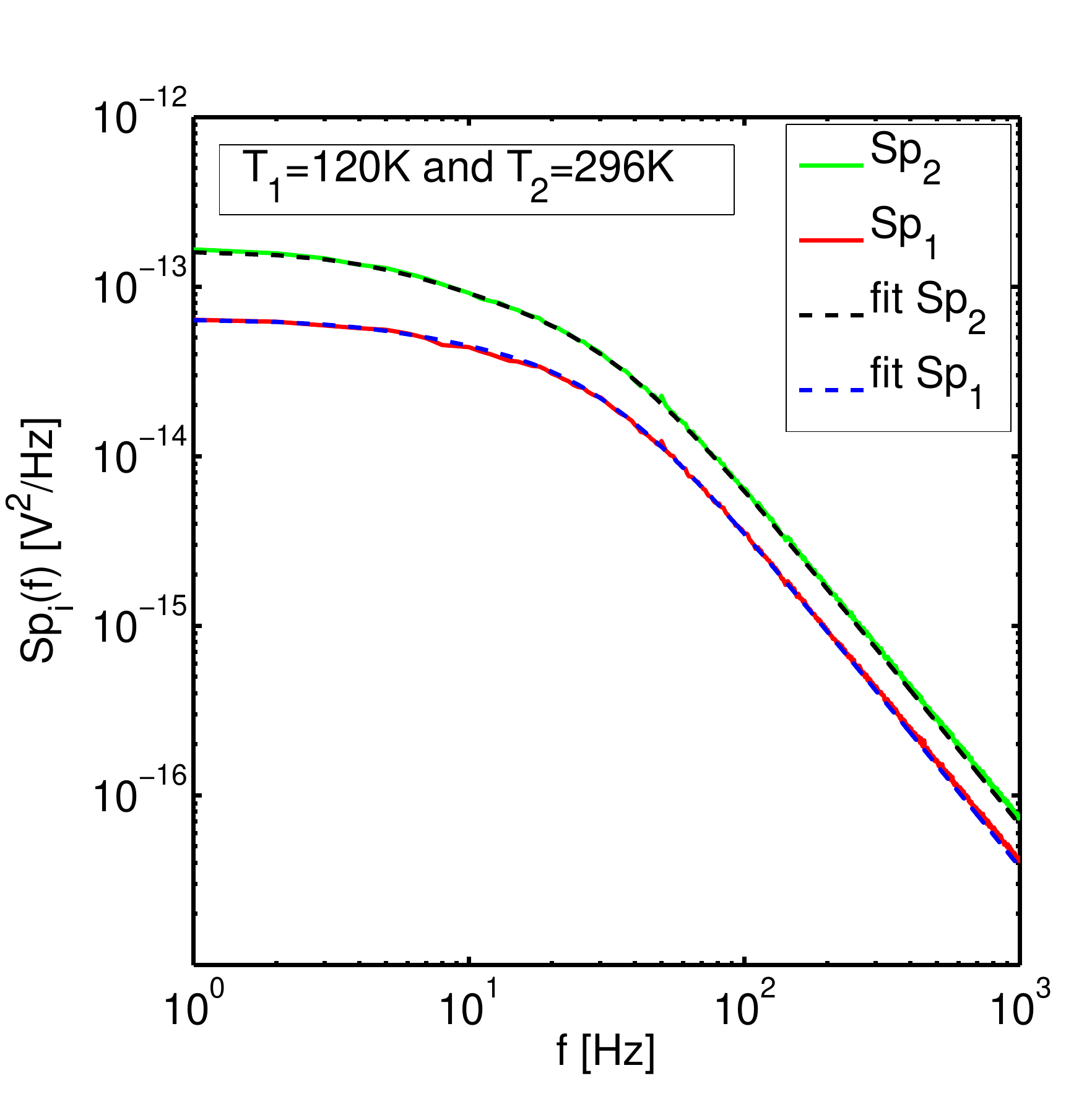}
   \includegraphics[width=0.3\textwidth]{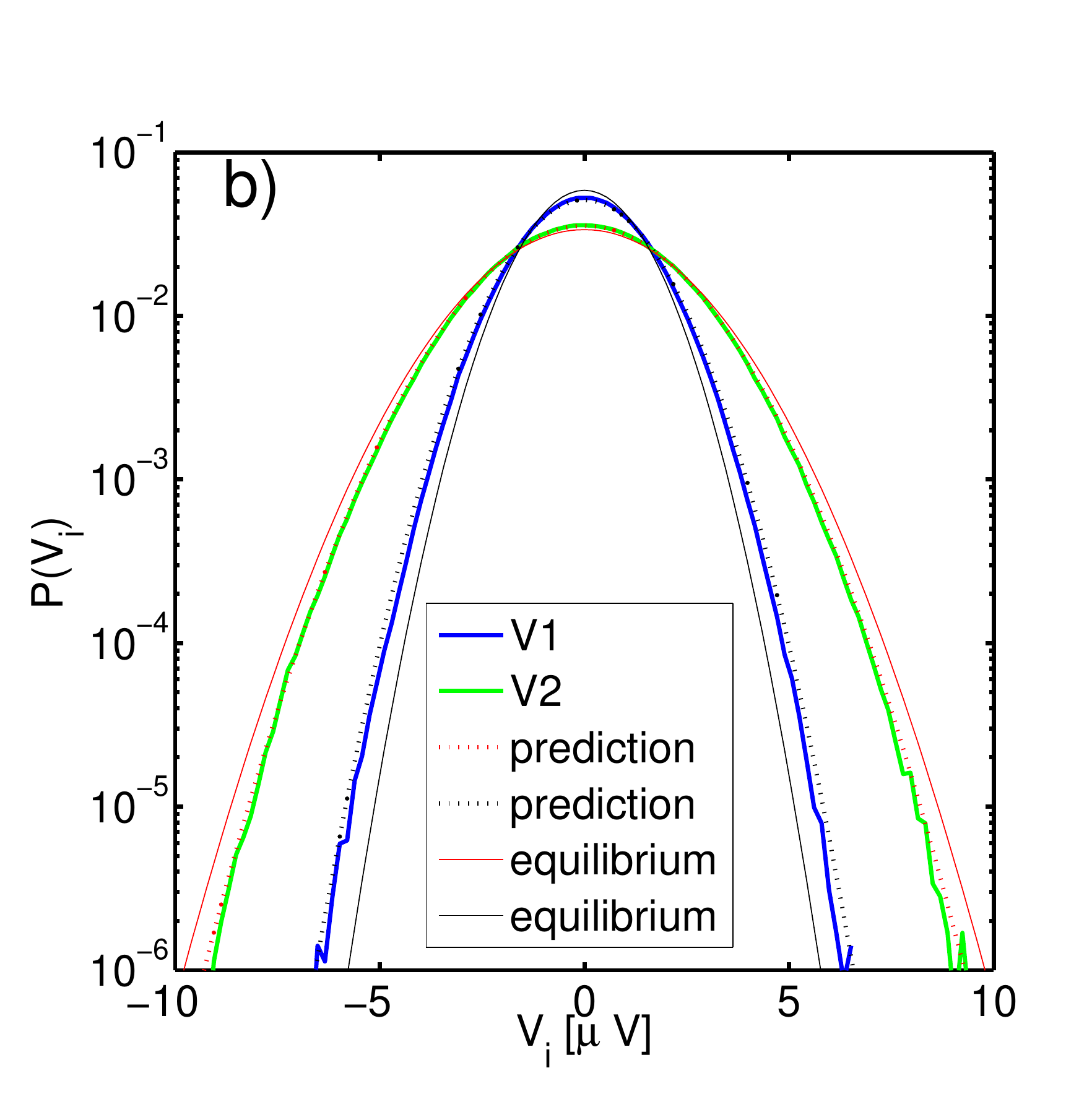}
     \caption{ 
  a)   The power spectra $Sp_1$ of  $V_1$  and $Sp_2$ of  $V_2$  measured  at $T_1=120K$  and $T_2=296K$ ($C=100pF,C_1=680pF,C_2=430pF$) are compared with the prediction of eq.~(\ref{eq.spectra_out1}) and (\ref{eq.spectra_out2}) (dashed lines)   
     The measured PDF of $V_1$ and $V_2$  are compared with the theoretical predictions in equilibrium and out of equilibrium obtained using the variance computed from eq.~(\ref{eq:s1_Q}).     
   b)  The corresponding Probability Density Function $P(V_1)$ of $V_1$  (green line) and  $P(V_2)$ of $V_2$ (blue line) measured  at $T_1=120K$ and $T_2=296K$. Dotted lines are the out-of-equilibrium PDF, whose variance  is estimated from the measure of the heat flux  (see fig.\ref{fig:P_Q}) and  eq.\ref{eq:s1_Q}. The continuous red line is the equilibrium $P(V_2)$ at $T_1=T_2=296K$ and  the black continuous line  corresponds to the equilibrium $P(V_1)$ at $T_1=T_2=120K$.   }
     \label{fig:S-spectre}
\end{figure}

These equations have been obtained by Fourier transforming the stochastic equations for the voltages eqs.~(\ref{lanV1})--(\ref{lanV2}), solving for $\tilde V_1(\omega)$ and $\tilde V_2(\omega) $ and computing the modula. 
The integral of eqs.~(\ref{eq.spectra_out1})  and (\ref{eq.spectra_out2}) gives the variances of $V_m$ (as given by eq.~(\ref{eq:s1})-(\ref{eq:s2})) directly computed from the distributions. Notice that the spectra eqs.~(\ref{eq.spectra_out1}) and (\ref{eq.spectra_out2}) contains the equilibrium parts given by 
eqs.~(\ref{eq:equilibrium_spectra_1})
 and (\ref{eq:equilibrium_spectra_2})  and an out of equilibrium component proportional to the temperature difference.
A comparison of eqs.~(\ref{eq.spectra_out1})--(\ref{eq.spectra_out2}) to the experimental power spectra is shown in fig.~\ref{fig:S-spectre}a).
In fig.~\ref{fig:S-spectre}b) we compare the measured probability distribution function (PDF)  of $V_1$  and $V_2$ with the  equilibrium and the out-of-equilibrium  distributions as computed by using  the theoretical predictions eqs.~(\ref{eq:s1})--(\ref{eq:s2}) for the variance.

\begin{figure}[h]
     \centering
       \includegraphics[width=0.22\textwidth]{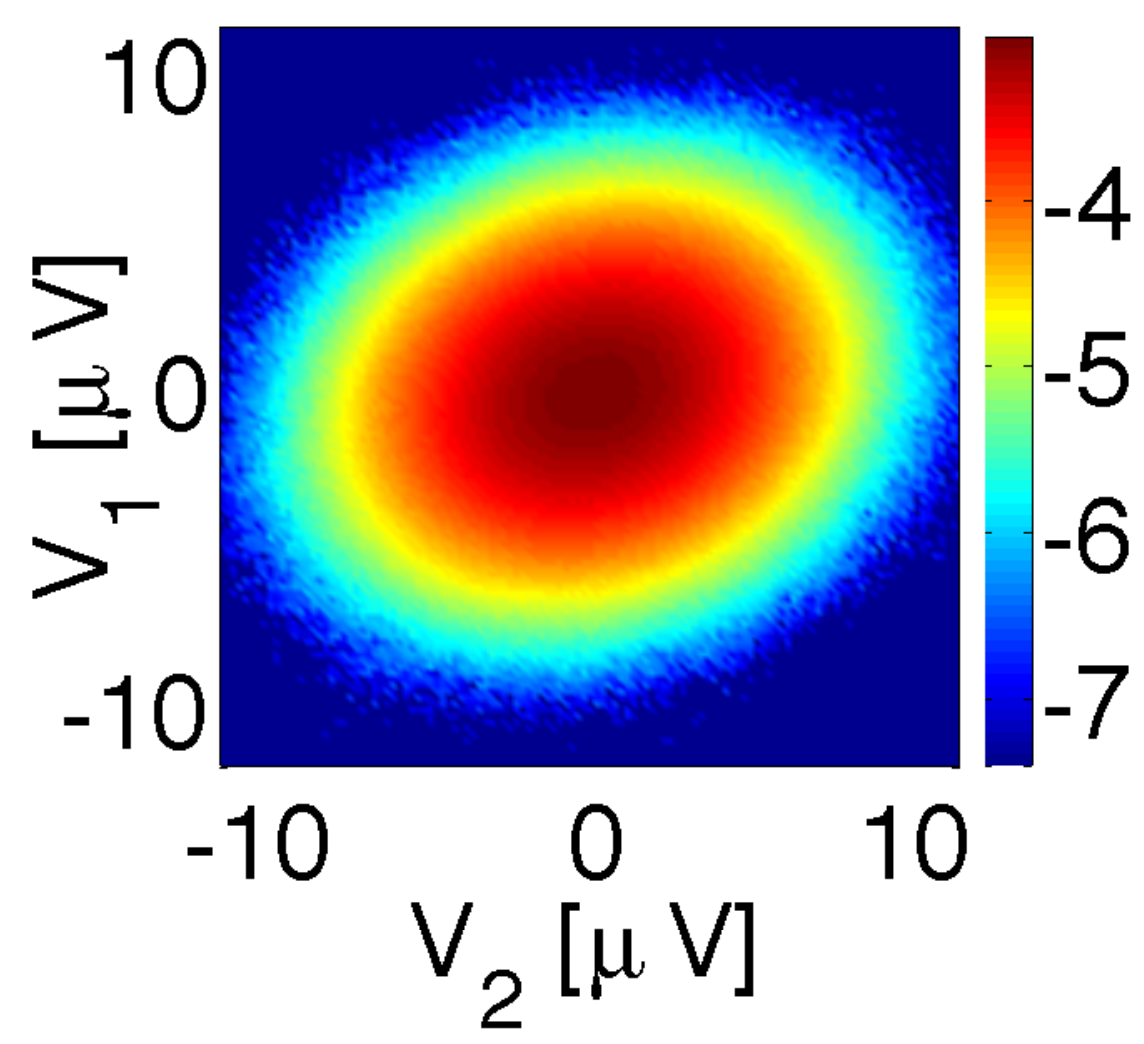}
    \includegraphics[width=0.22\textwidth]{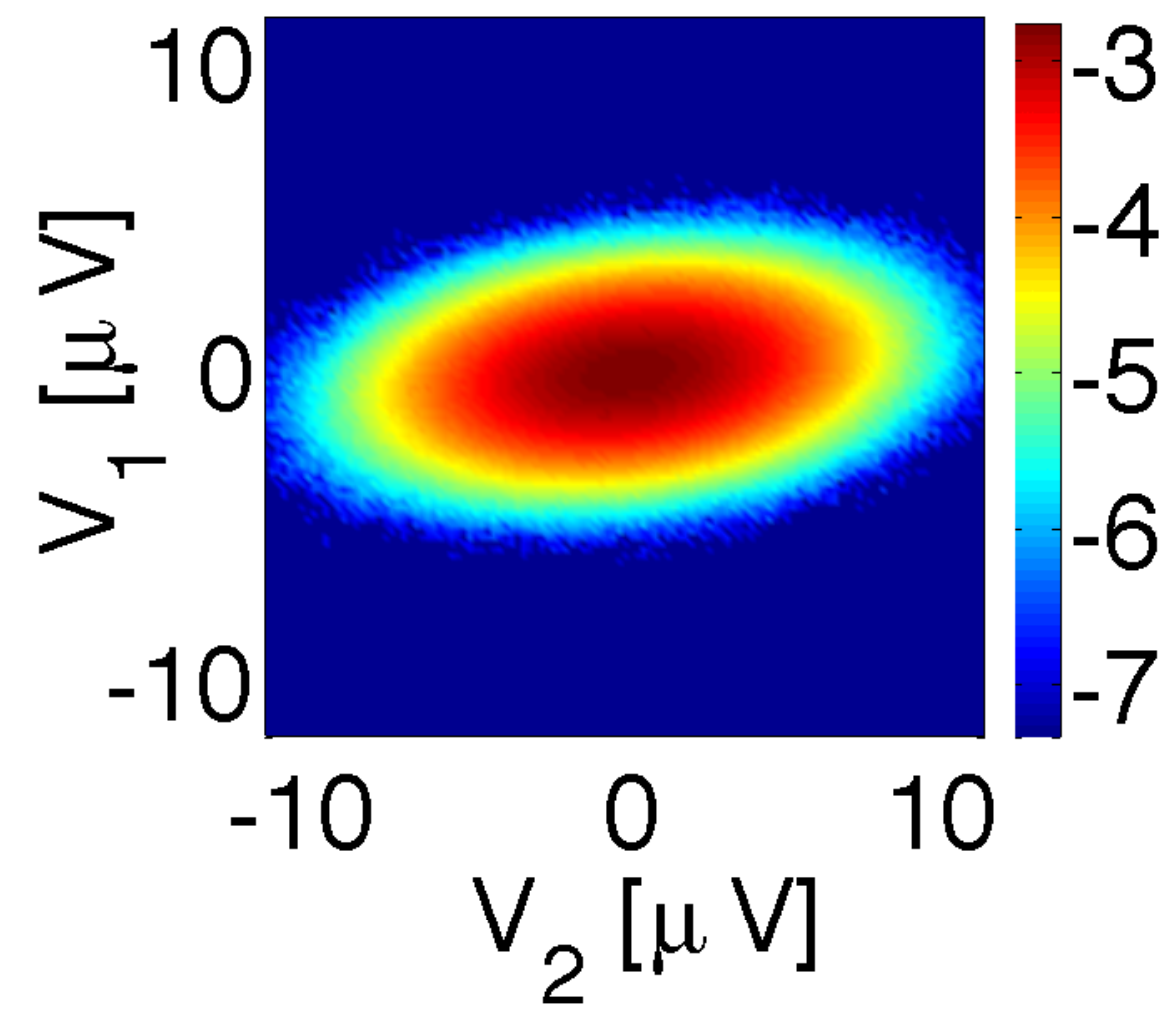}
     \caption{The joint probability $\log_{10}P(V_1,V_2)$  measured at $T_1=296K$ equilibrium (a) and out of equilibrium $T_1=88K$(b).
     The color scale   is indicated on the colorbar on the right side. 
      }
     \label{fig:pdfV1V2}
\end{figure}

\subsubsection{ The joint probability of $V_1$ and $V_2$}
As discussed in sections \ref{stoch:sec} and \ref{fluct:sec}, all the relevant thermodynamic quantities can be sampled once one has measured the  voltage across the resistors $V_1$, $V_2$. The fluctuations of these quantities are thus to be fully characterized before one can proceed and study the fluctuations of all the derived thermodynamic quantities.
Thus, we first study the joint probability distribution $P(V_1,V_2)$, which is plotted in fig.~\ref{fig:pdfV1V2}a) for  $T_1=T_2$ and in 
fig.~\ref{fig:pdfV1V2}b) for   $T_1=88K$. The fact that the axis of the  ellipses defining the contours lines of  $P(V_1,V_2)$ are inclined with respect to the $x$ and $y$ axis indicates that there is a certain correlation between $V_1$ and $V_2$.  This correlation, produced by the electric coupling, plays a  major role in determining the mean  heat flux between the two reservoirs, as we discuss below. 
We are mainly interested in the out-of-equilibrium case, when $T_1\ne T_2$, and in the following, we will characterize 
 the heat flux and the entropy production rate, and discuss how the variance of $V_1$ an $V_2$ are modified by the presence of a non-zero heat flux.

\subsection{Heat flux fluctuations} \label{section_Q_sigma}
\begin{figure}[h]
     \centering
    \includegraphics[width=0.23\textwidth]{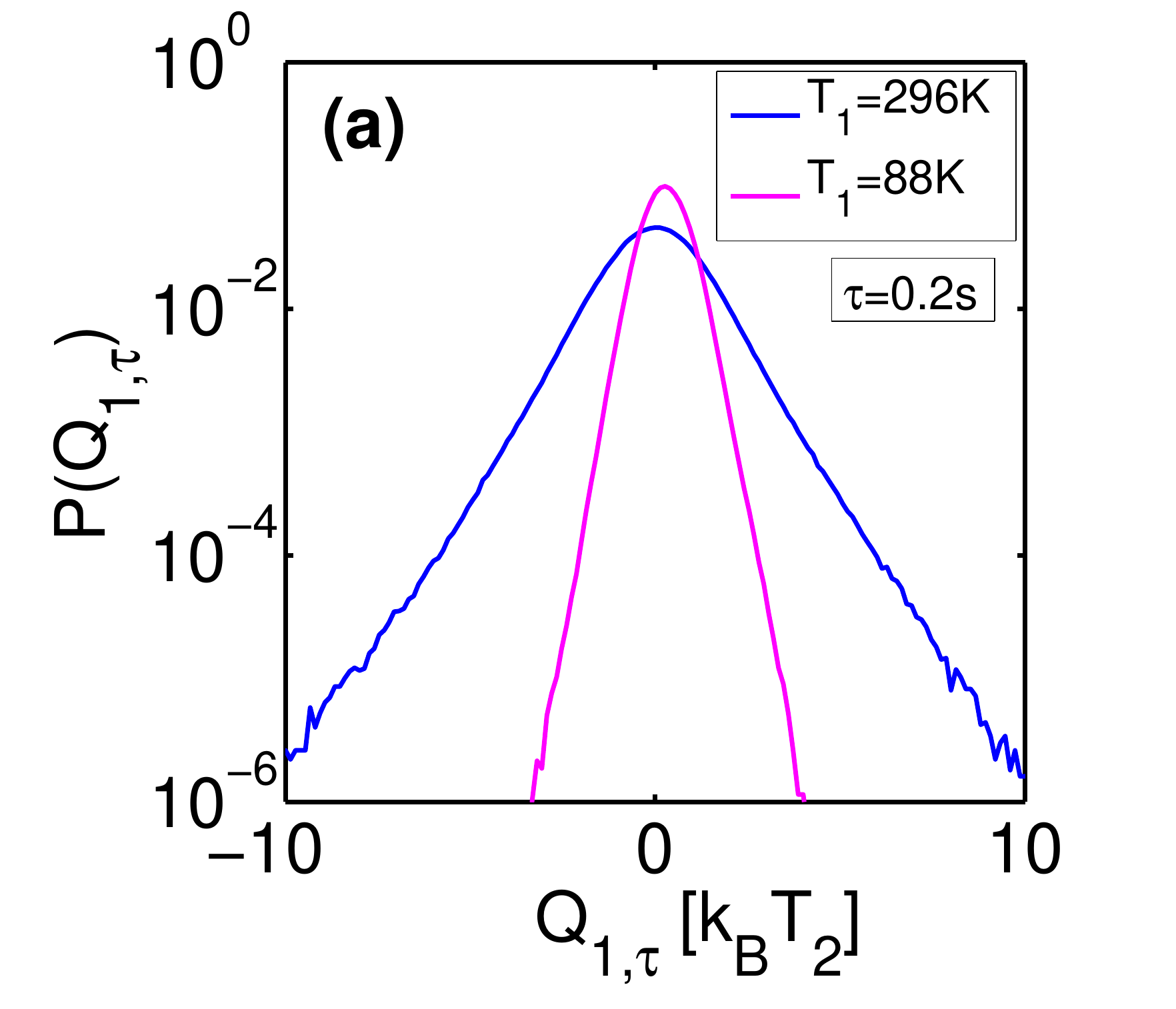} 
    \includegraphics[width=0.23\textwidth]{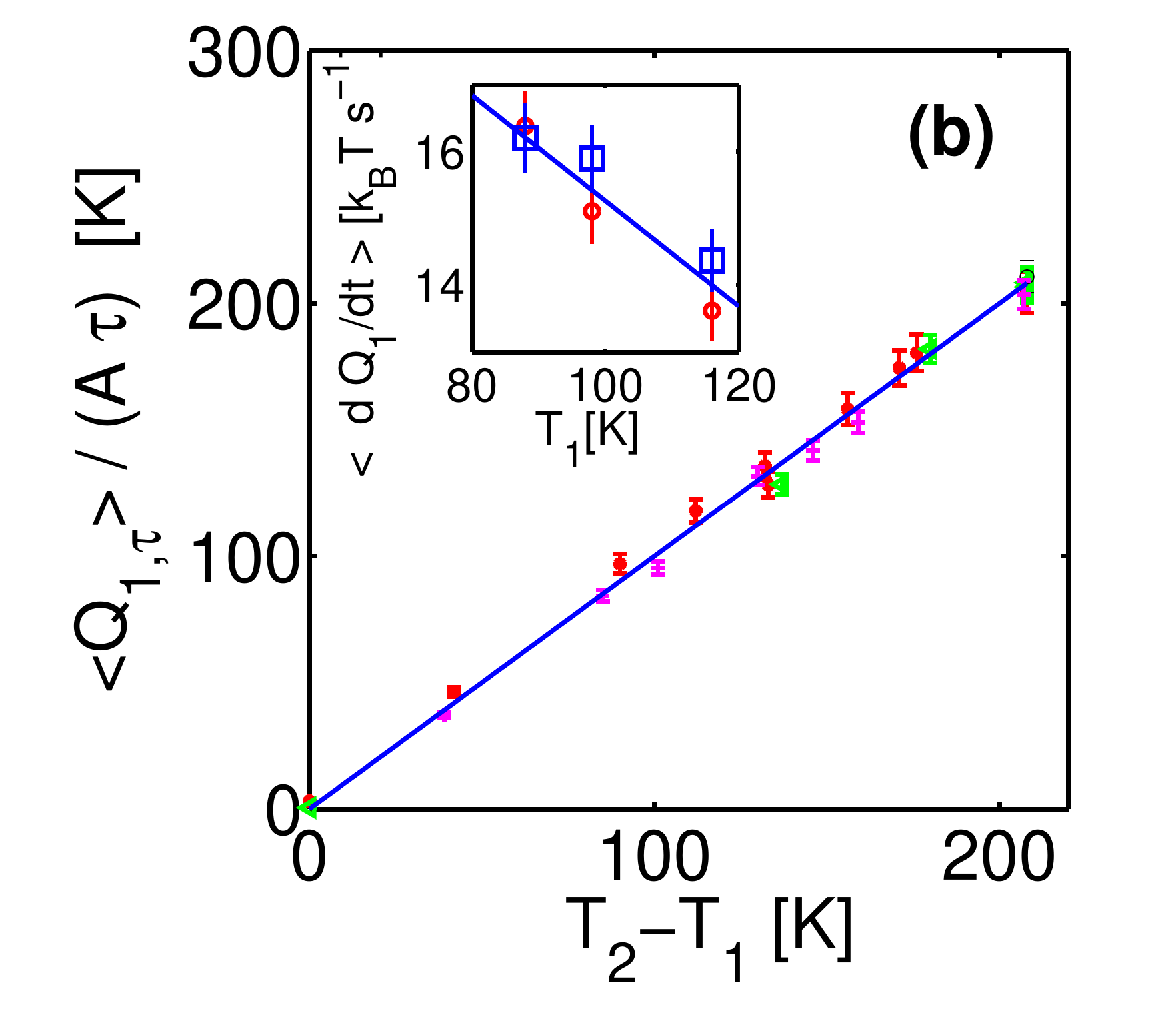}
     \caption{a) The  probability $P(Q_{1,\tau})$  measured at $T_1=296K$ (blue line) equilibrium 
    and  $T_1=88K$ (magenta line) out of equilibrium. Notice that the peak of the $P(Q_{1,\tau})$ is centered at zero at equilibrium and shifted towards a positive value out of equilibrium. The amount of the shift is very small and  {is $\sim k_B(T_2-T_1)$.}   b) The measured mean value of $\average{Q_{1,\tau}}$ is a linear  function of $(T_2-T_1)$.  The red points correspond to measurements performed with the values of the capacitance  $C_1,C_2,C$ given in the text and $\tau=0.2s$.  The other  symbols and colors pertain to different values of these capacitance and other  $\tau$: {(black $\circ$) $\tau=0.4s,C=1000pF$, (green $\triangleleft$) $\tau=0.1s,C=100pF$, (magenta $+$) $\tau=0.5s,C=100pF$}.  The values of $\average{Q_{1,\tau}}$ have been rescaled by the parameter dependent theoretical prefactor $A$, which allows the comparison of different experimental configurations. The continuous blue line with slope $1$ is the theoretical prediction of eq.~\ref{dtQ1}. 
    In the inset the values of $<\dot Q_1>$ (at $C=1000pF$) directly measured using $P(Q_1)$ (blue square) are compared with those (red circles)  obtained  from the eq. (\ref{eq:s1_Q}). 
}
\label{fig:P_Q}
\end{figure}

In fig.~\ref{fig:P_Q}a) we show the probability density function  $P(Q_{1,\tau})$, at various temperatures:
we see that $Q_{1,\tau}$ is a strongly fluctuating quantity, whose PDF $P(Q_{1,\tau})$ has long exponential tails.
Notice that although for $T_1< T_2$ the mean value of $Q_{1,\tau}$ is positive, instantaneous negative fluctuations may occur,  i.e., sometimes the heat flux is reversed.   The mean values of the dissipated heat are expected to be  linear functions of  the temperature difference $\Delta T=T_2-T_1$, i.e. $\average{Q_{1,\tau}}=A \,  \tau \,\Delta T$, where $A=k_B C^2/XY$ is a parameter dependent quantity, that can be obtained by eq.~(\ref{dtQ1}). 
This relation is confirmed by our experimental results, as shown in fig.~\ref{fig:P_Q}b. Furthermore, the mean values of the dissipated heat satisfy the equality $\average{Q_2}=-\average{Q_1} $, corresponding to an energy conservation principle: the power extracted from 
the bath 2 is dissipated into the bath 1 because of the electric coupling.

As we discuss in section \ref{secFTQ1}, the mean heat flow is related to a change in the variances 
$\sigma_m^2(T_m)$ of  
$V_m$  with respect to the equilibrium value  $\smeq^2(T_m)$, see eq.~(\ref{eq:s1_Q}).
The experimental verification of eq.~(\ref{eq:s1_Q}) is shown in the inset of fig.~\ref{fig:P_Q}b) where the values of $\average{\dot Q_m}$ directly estimated from the experimental data (using the steady state $P(Q_m)$) are compared with those  obtained from the difference of the variances of  $V_1$ measured in equilibrium and out-of-equilibrium. The values are comparable within the error bars and show that the out-of-equilibrium variances  are modified only by the heat flux.  

\subsection{Fluctuation theorem for  work and heat } \label{FT:sec}

\begin{figure}[h]
     \centering
       \includegraphics[width=0.23\textwidth]{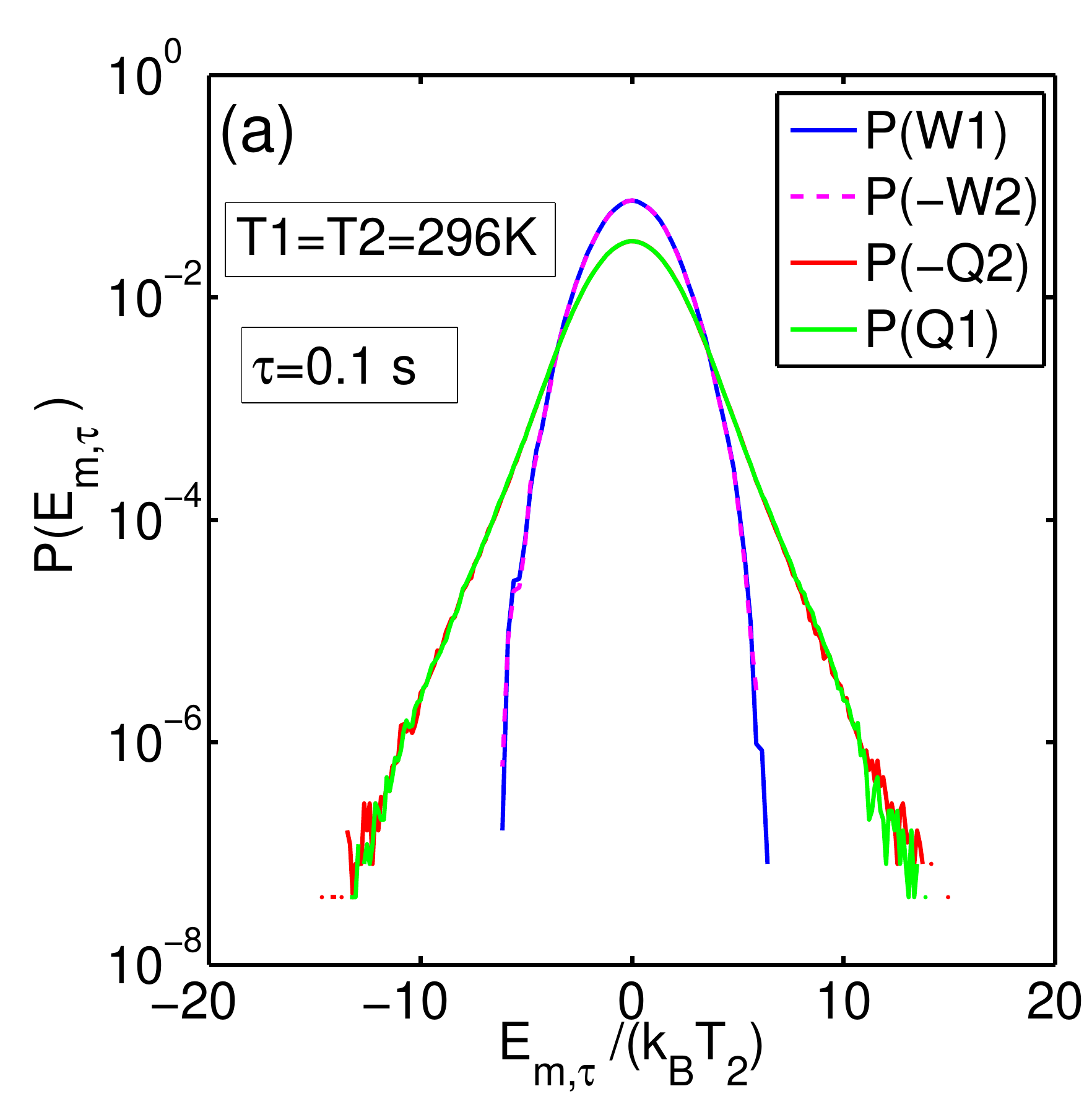}  
        \includegraphics[width=0.23\textwidth]{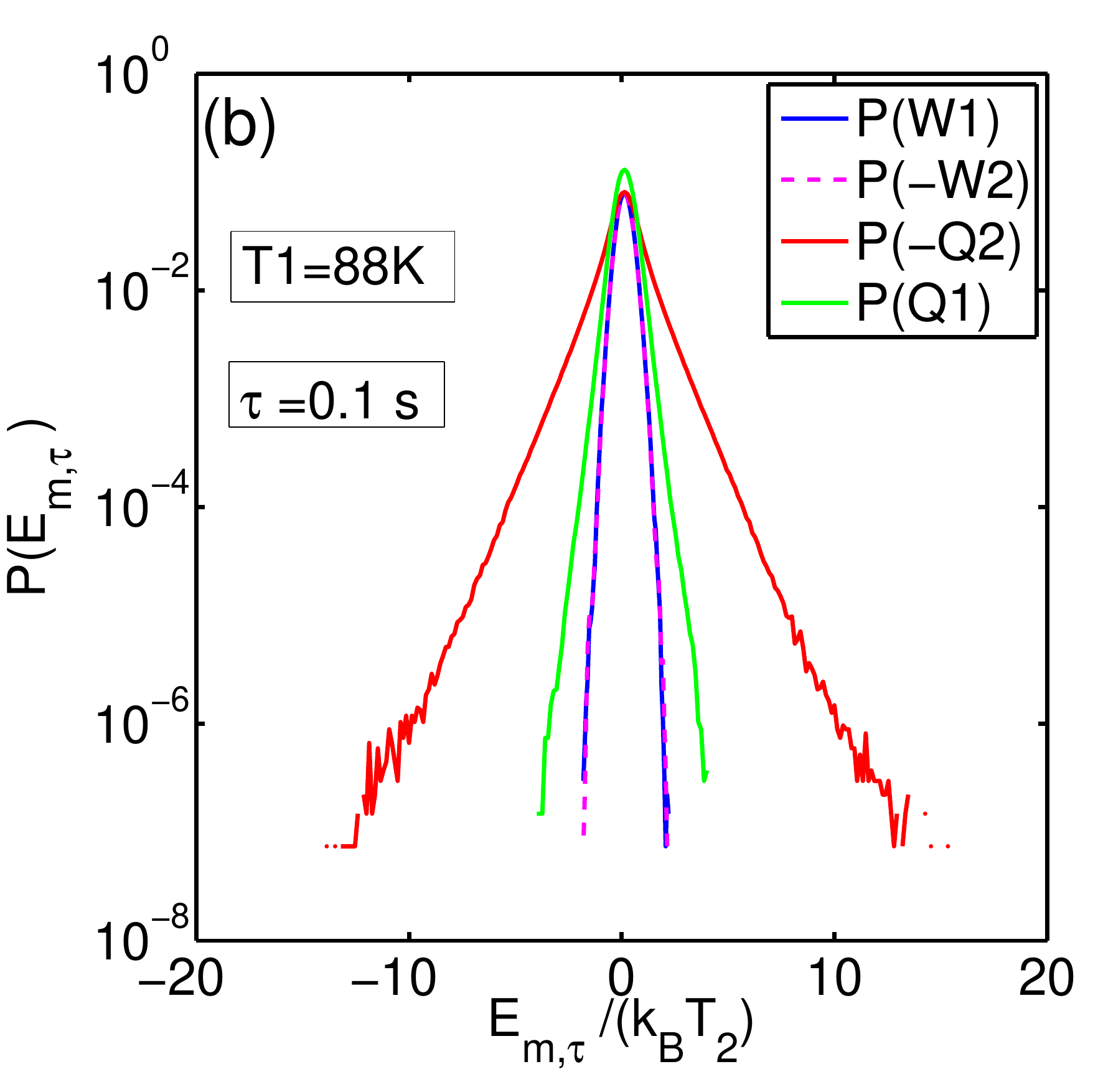} \\
         \includegraphics[width=0.23\textwidth]{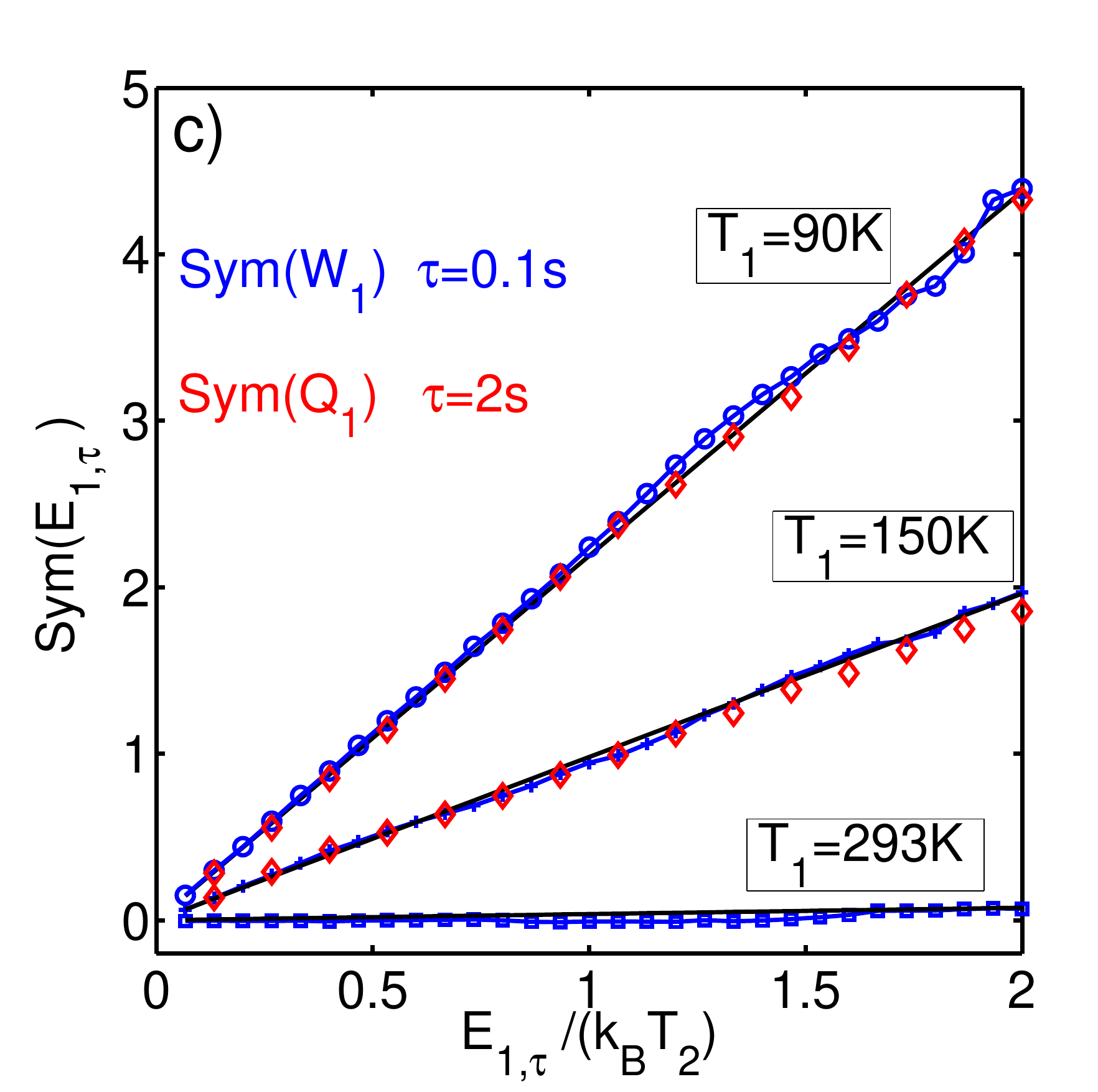}
     \includegraphics[width=0.23\textwidth]{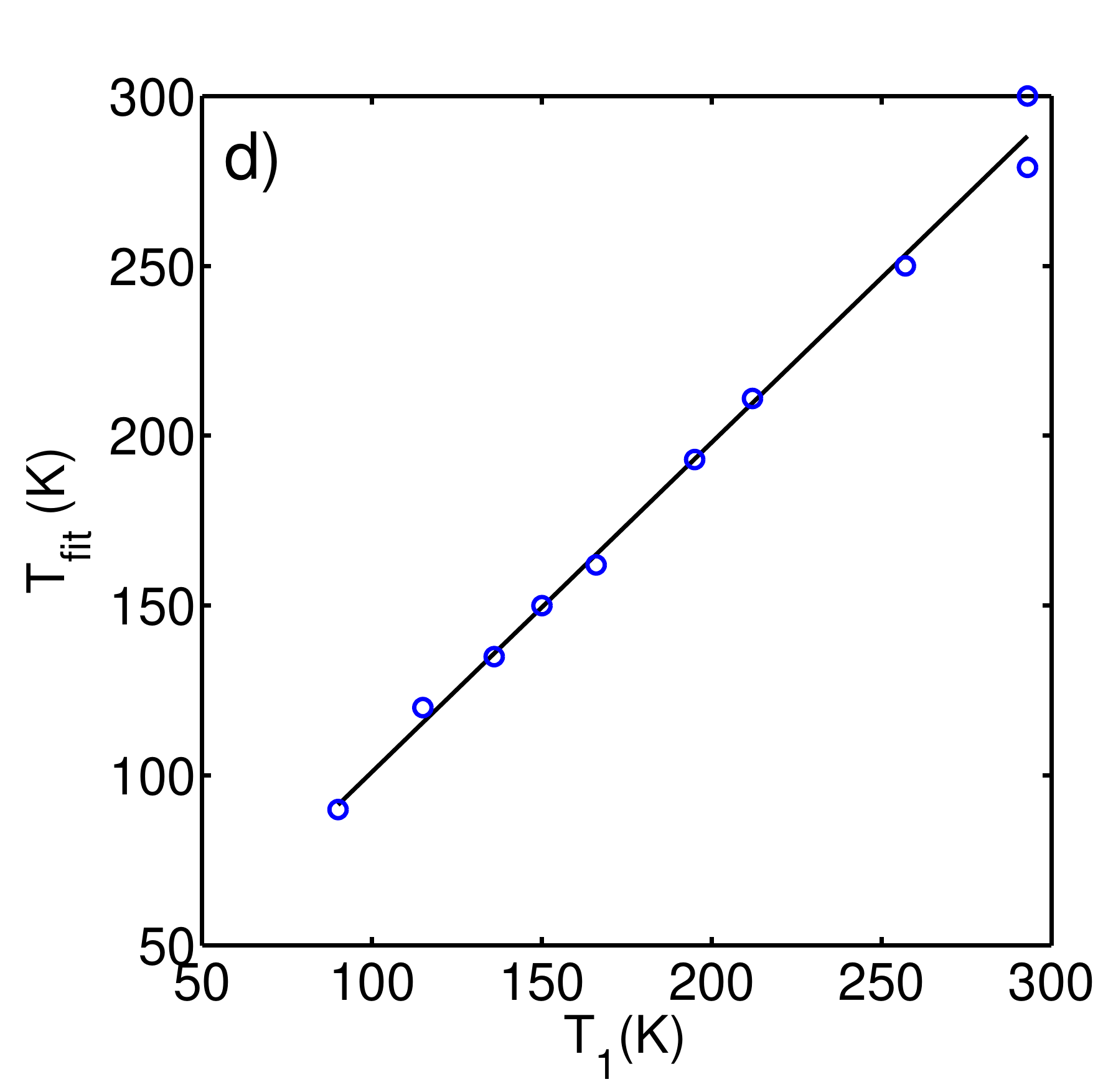}
     \caption{ a)  Equilibrium: $P(W_{m,\tau})$ and $P(Q_{m,\tau})$,  measured in equilibrium at $T_1=T_2=296K$ and $\tau=0.1s$, are plotted as functions of  $E$, where $E$ stands either for $W$ or $Q$. Notice that, being the system in equilibrium $P(W_{1,\tau}) =P(-W_{2,\tau})$, $P(Q_{1,\tau}) =P(-Q_{2,\tau})$.  b) Out of equilibrium: same distributions as in  a) but the PDFs are measured at $T_1=88K$, $T_2=296K$ and $\tau=0.1s$. Notice that in out of equilibrium $P(W_{1,\tau}) =P(-W_{2,\tau})$ but $P(Q_{1,\tau}) \ne P(-Q_{2,\tau})$. The reason of this difference is explained in the text. c) The symmetry function $Sym(E_{1,\tau})$, measured a various $T_1$ is plotted as a function of $E_1$ ($W_1$ or $Q_1$). The theoretical slope of these straight lines is $T_2/T_1-1$. d) The temperature $T_{fit}$ estimated from the slopes of the lines in c) is plotted  as a function of the $T_1$ measured by the thermometer.  The slope of the line is 1 showing that $T_{fit}\simeq T_1 $ within a few percent.     }
\label{fig:P_X}
\end{figure}

As the system is in a stationary state,  we have $\average{W_{m,\tau}}=\average{Q_{\tau,m}}$. Instead the comparison of the pdf of $W_{m,\tau}$ with those of $Q_{\tau,m}$, measured  at various temperatures, presents several interesting features. In fig. \ref{fig:P_X}(a) we plot $P(W_{1,\tau})$, $P(-W_{2,\tau})$, $P(Q_{1,\tau})$ and $P(-Q_{2,\tau})$ measured in equilibrium at $T_1=T_2=296K$ and $\tau\simeq 0.1 s \simeq 10 \, Y$. We immediately see that the fluctuations of the work are almost Gaussian whereas those of the heat presents large exponential tails. 
This well known difference \cite{VanZonCil} between  $P(Q_{m,\tau})$ and $P(W{m,\tau})$ is induced by the fact that $Q_{m,\tau}$ depends also on $\Delta U_{m,\tau}$ (eq.\ref{eq:DeltaUm}), which is the sum of the square of Gaussian distributed variables, thus inducing exponential tails in  $P(Q_{m,\tau})$. In fig. \ref{fig:P_X}(a) we also  notice that
 $P(W_{1,\tau})=P(-W_{2,\tau})$ 
and  $P(Q_{1,\tau})=P(-Q_{2,\tau})$, showing that in equilibrium all fluctuations are perfectly symmetric. The same pdfs measured in the out of equilibrium case at $T_1=88K$  are plotted in fig. \ref{fig:P_X}(b). We notice here that in this case  the behavior of the pdfs of the heat is different from those of the work. Indeed although  $\average{W_{m,\tau}}>0$ we observe that $P(W_{1,\tau})=P(-W_{2,\tau})$, while 
$P(Q_{1,\tau})\ne P(-Q_{2,\tau})$. Indeed the shape  of  $P(Q_{1,\tau})$ is strongly modified by changing $T_1$ from $296K$ to 
$88K$, whereas the shape of  $P(-Q_{2,\tau})$ is slightly modified   by the large temperature change, only the tails of  $P(-Q_{2,\tau})$ presents a small asymmetry testifying  the presence of a small heat flux. 
The fact that  $P(Q_{1,\tau})\ne P(-Q_{2,\tau})$ whereas $P(W_{1,\tau})= P(-W_{2,\tau})$ can be understood by noticing that $Q_{m,\tau}=W_{m,\tau}-\Delta U_{m,\tau}$.  Indeed $\Delta U_{m,\tau}$ (eq.\ref{eq:DeltaUm}) depends on the values of $C_m$ and $V_m^2$. As  $C_1\ne C_2$ and $\sigma_2 \ge \sigma_1$, this explains the different behavior of $Q_1$ and $Q_2$. Instead $W_m$  depends only on $C$ and the product $V_1 \, V_2$. 

 We have studied whether our data satisfy the fluctuation theorem as given by eq.~(\ref{eq_Pq}) in the limit of large $\tau$. It turns out that the symmetry imposed by eq.~(\ref{eq_Pq}) is reached for rather small $\tau$ for $W$. Instead it converges very slowly for $Q$.   We only have a qualitative argument to explain this difference in the asymptotic  behavior: by looking at the data one understands that the slow convergence is induced by the presence of the exponential tails of $P(Q_{1,\tau})$ for small $\tau$. 
  
To check eq.~\ref{eq_Pq}, we plot in fig.~\ref{fig:P_X}c) the symmetry function 
$Sym(E_{1,\tau})=\ln {P(E_{1,\tau}) \over P(-E_{1,\tau})} $ as a function of $E_{1,\tau}/(k_BT_2)$ measured at different $T_1$, but $\tau=0.1s$ for $Sym(W_{1,\tau})$ and $\tau=2s\simeq 200Y$ for $Sym(Q_{1,\tau})$. Indeed for  $Sym(Q_{1,\tau})$ reaches the asymptotic regime only for $\tau<2s$. We see that  $Sym(W_{1,\tau})$ is a linear function  of $W_{1,\tau}/(k_BT_2)$ at all $T_1$. These straight lines have a slope  $\alpha(T_1)$ which, according to  eq.\ref{eq_Pq}  should be  $(\beta_{12} k_B T_2)$. In order to check this  prediction we fit the slopes of the 
straight lines in fig.\ref{eq_Pq}c). From the fitted $\alpha(T_1)$ we deduce a temperature $T_{fit}= T_2/(\alpha(T1)+1)$ which is compared to the measured temperature $T_1$ in fig.\ref{eq_Pq}d). In this figure the straight line of slope 1 indicates that $T_{fit}\simeq T_1$ within a few percent. These experimental results indicate that our data verify the fluctuation theorem, eq.\ref{eq_Pq},  for the work and the heat but that the asymptotic regime is reached for much larger time for the latter. 

\subsection{Statistical properties of entropy}
\begin{figure}[h]
     \centering
    \includegraphics[width=0.238\textwidth]{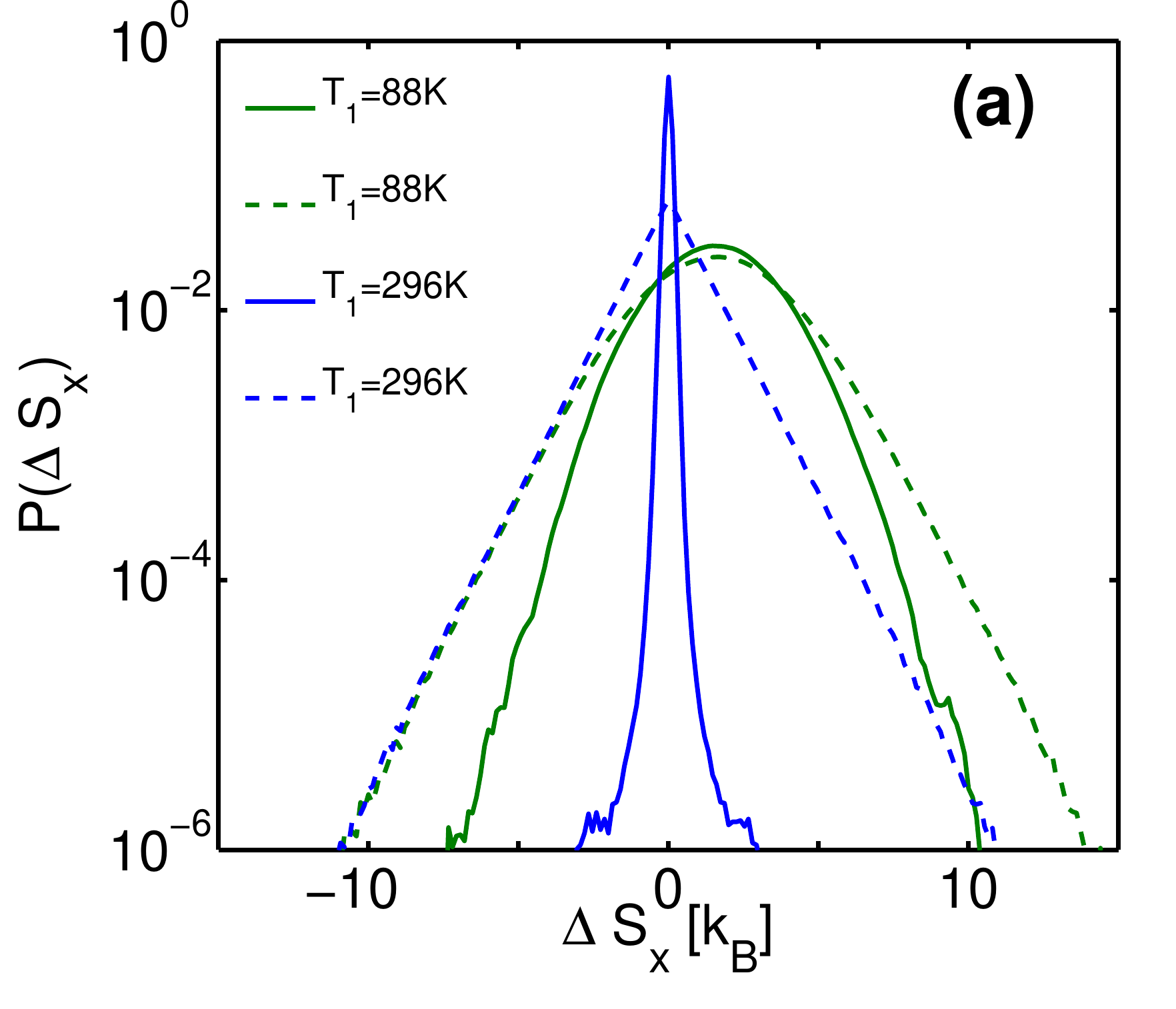}
    \includegraphics[width=0.238\textwidth]{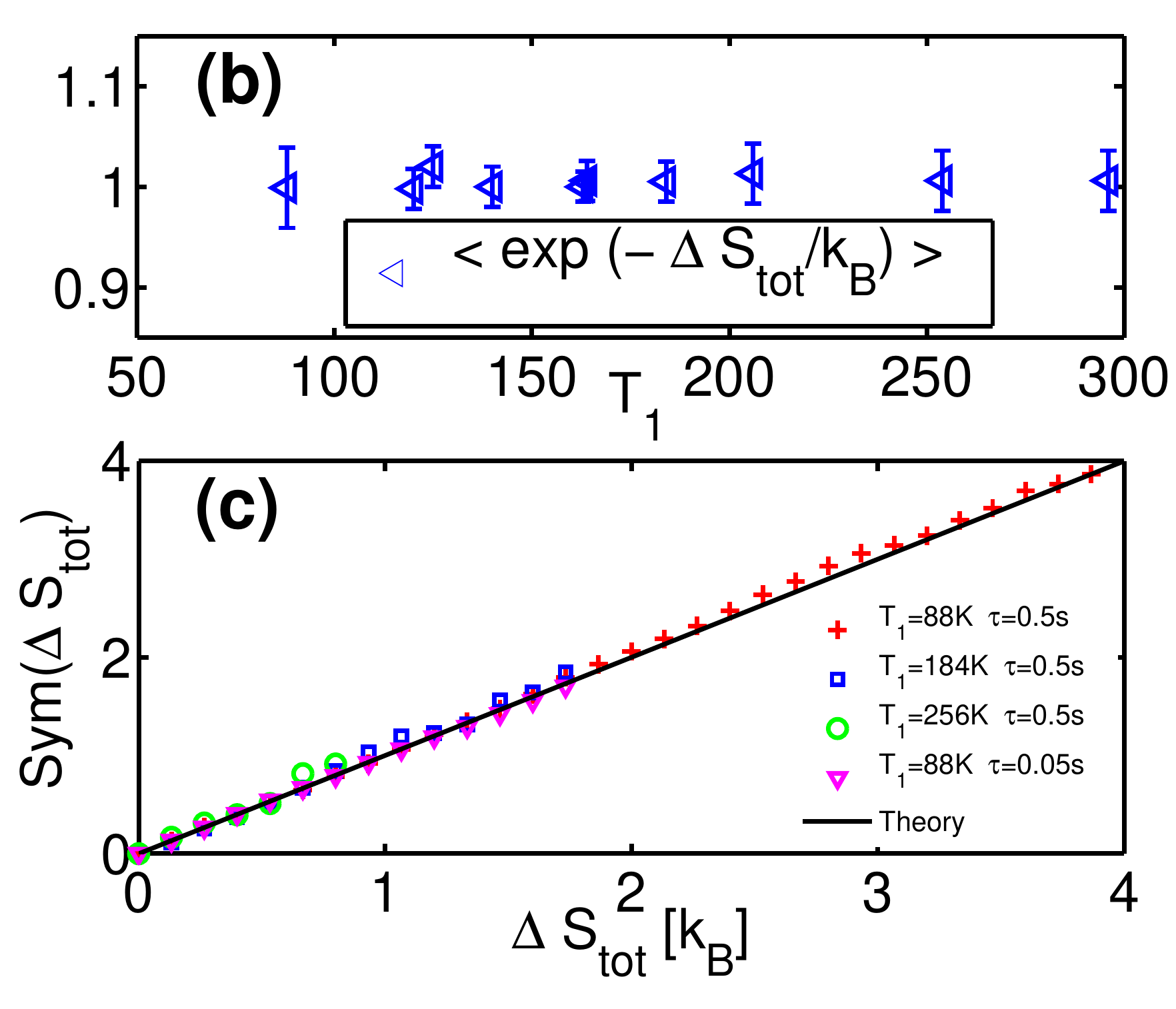}
     \caption{ a) The  probability $P(\Delta S_r)$ (dashed lines) and $P(\Delta S_{tot})$ (continuous lines)   measured at $T_1=296K$ (blue line) which corresponds to equilibrium 
    and  $T_1=88K$ (green lines) out of equilibrium. Notice that both distributions are  centered at zero at equilibrium and shifted towards positive value in the out-of-equilibrium.    b)  $\average{\exp(-\Delta S_{tot})}$ as a function of $T_1$ at two different $\tau=0.5s$ and $\tau=0.1s$.  c) Symmetry function $\rm{Sym}(\Delta S_{tot})=\log[P(\Delta S_{tot})/P(-\Delta S_{tot})]$ as a function of $\Delta S_{tot}$. The black straight line of slope 1 corresponds to the theoretical prediction. }
     \label{fig:P_DS}
\end{figure}

We now turn our attention to the study of the entropy produced by the  total system, circuit plus heat reservoirs.  We consider first the entropy  $\Delta S_{r,\tau}$ due to the heat exchanged with the reservoirs, which reads  $\Delta S_{r,\tau}= Q_{1,\tau}/T_1 +Q_{2,\tau}/T_2$. This entropy is a fluctuating quantity as both $Q_1$ and  $Q_2$ fluctuate, and its  average in a time $\tau$ is 
$\average{\Delta S_{r,\tau}} = \average{Q_{r,\tau}}(1/T_1-1/T_2)=A \tau (T_2-T_1)^2 /(T_2 \, T_1)$.   However the reservoir entropy $\Delta S_{r,\tau}$ is not the only component of the total entropy production: one has to take into account the  entropy variation of the system, due to its dynamical evolution.
Indeed, the state variables $V_m$  also fluctuate as an effect of the thermal noise, and thus, if one measures their values at regular time interval, one obtains a ``trajectory'' in the phase space $(V_1(t), V_2(t))$. Thus, following Seifert \cite{Seifert}, who developed this concept for a single heat bath, one can introduce a trajectory  entropy for the evolving system $S_s(t)=-k_B \log P (V_1(t), V_2(t))$, which extends to non-equilibrium systems the standard Gibbs entropy concept.  Therefore, when evaluating the total entropy production, one has to take into account the contribution over the time interval $\tau$ of 
\begin{equation}
\Delta S_{s,\tau}= - k_B  \log \left[{ P(V_1(t+\tau ),V_2(t+\tau )) \over P(V_1(t),V_2(t))} \right].
\label{eq:DS_tot}
\end{equation}
It is worth noting that the system we consider is in a non-equilibrium steady state, with a constant external driving $\Delta T$. Therefore the probability distribution $P(V_1,V_2)$ (as shown in fig.~\ref{fig:pdfV1V2}b)) does not depend explicitly on the time, and   $\Delta S_{s,\tau}$
is non vanishing whenever the final point of the trajectory is different
from the initial one: $(V_1(t+\tau ),V_2(t+\tau ))\neq (V_1(t),V_2(t))$.
Thus the total entropy {change} reads $\Delta S_{tot,\tau}=\Delta  S_{r,\tau} +\Delta S_{s,\tau}$, where we omit the explicit dependence on $t$, as the system is in a steady-state as discussed above. This entropy has  several interesting features. The first  one is  that $\average{\Delta S_{s,\tau}}=0 $, and as a consequence $\average{\Delta S_{tot}}=\average{ \Delta S_r}$  which grows with increasing  $\Delta T$. The second and
 most interesting result is that independently of $\Delta T$ and of $\tau$, the following equality always holds: 
\begin{equation}
\average{\exp(-\Delta S_{tot} /k_B)}=1,
\label{eq:DS}
\end{equation}
for which we find both experimental evidence, as discussed in the following, and provide a theoretical proof in appendix \ref{app}. Equation~(\ref{eq:DS}) represents an extension to two temperature sources of the result obtained for  a system in a single heat bath driven out-of-equilibrium by a time dependent mechanical force~\cite{Seifert,Evans02} and our results provide the first experimental verification of the expression in  a system driven by a temperature difference. Eq.~(\ref{eq:DS}) implies that $\average{\Delta S_{tot} } \, \ge  0$, as prescribed by the second law.  {From symmetry considerations, it follows immediately that, at equilibrium ($T_1=T_2$), the probability distribution of $\Delta S_{tot}$ is symmetric: $P_{eq}(\Delta S_{tot})=P_{eq}(-\Delta S_{tot})$. Thus Eq.~(\ref{eq:DS}) implies  that  the probability density function of $\Delta S_{tot}$ is a Dirac $\delta$ function  when $T_1=T_2$, i.e. the quantity $\Delta S_{tot}$ is rigorously zero in equilibrium, both in average and fluctuations, and so its mean value  and  variance provide a measure of the entropy production.}
{The  measured probabilities $P(\Delta S_r)$ and $P(\Delta S_{tot})$ are  shown in fig. \ref{fig:P_DS}a). We see that  $P(\Delta S_{r})$  and
 $P(\Delta S_{tot})$ are quite different and that the latter is close to a Gaussian and reduces to a Dirac $\delta$ function in equilibrium, i.e. $T_1=T_2=296K$ (notice that, in fig.\ref{fig:P_DS}a,  the small broadening of the equilibrium $P(\Delta S_{tot})$ is just due to unavoidable experimental noise and  discretization of the experimental probability density functions). }The experimental measurements satisfy eq.~(\ref{eq:DS}) as it is shown in  fig. \ref{fig:P_DS}b). It is worth to note that eq.~(\ref{eq:DS}) implies that  $P(\Delta S_{tot})$ should satisfy a fluctuation theorem of the form
    $\log [P(\Delta S_{tot})/P(-\Delta S_{tot})]= \Delta S_{tot}/k_B, \, \, \,  \forall \tau,\Delta T $, as discussed extensively in reference \cite{Seifert_2012,EVDB10}.  We clearly see in fig.\ref{fig:P_DS}c) that this relation holds for different values of the temperature gradient.    
Thus this experiment clearly establishes  a relationship between the mean and the variance of the entropy production rate 
in a system driven out-of-equilibrium  by the temperature difference between two thermal baths coupled by electrical noise. Because of the formal analogy with Brownian motion the results also apply to mechanical coupling as discussed in the following. 

\section{Conclusions}\label{concl:sec}

We have studied experimentally and theoretically the statistical properties of the energy exchanged between two  heat baths at different temperatures which are coupled by electric thermal noise. We have measured the heat flux, the thermodynamic  work and the total entropy, and shown that each of these quantities exhibits a FT, in particular we have shown the existence of a conservation law for entropy  which is not asymptotic in time.  Our results  hold in full generality since the electric system considered here is ruled by 
the same equations as for two Brownian particles, held at  different temperatures and mechanically coupled by a conservative potential. Therefore  these results set precise constraints on the energy exchanged between coupled nano and micro-systems held at different temperatures. 
Our system can be easily scaled to include more than two heat reservoirs, and more electric elements to mimic more complex dynamics in a system of   Brownian particles. 
We thus believe that our study can represent the basis  for further investigation in out-of-equilibrium physics.

\section*{Acknowledgments}
This work has been partially supported by the French
Embassy in Denmark through the French-Danish scientific
co-operation program, by ESF network Exploring the
Physics of Small Devices and by the ERC contract OUTEFLUCOP. AI gratefully acknowledges financial support from  the Danish Research Council (FNU) through the project "Manipulating small objects with light and heat".

\appendix
\section{Entropy conservation law}\label{app}

We now turn our attention to eq.~(2), in the main text, and provide a formal proof for it.
In the present appendix we provide a formal proof of eq.~(\ref{eq:DS}).
Let's divide the time into small intervals $\Delta t$, and  let  
$\bV=(V_1,V_2)$ denote the system's stat at time $t$, and $\bV'=(V_1+\Delta V_1,V_2+\Delta V_2)$ its state at time $t+\Delta t$. Let $\mathcal P_F(\bV\rightarrow \bV'| \bV,t)$ be the probability that the system undergoes a transition from $\bV$ to $\bV'$ provided that its state at time $t$ is $\bV$, and let  $\mathcal P_R(\bV'\rightarrow \bV| \bV',t+\Delta t)$ be the probability of the time-reverse transition.
By noticing that the time evolution of the dynamic variables $V_m$ is ruled by eqs.~(\ref{lanV1})-(\ref{lanV2}), we find that the probability of the forward trajectory can be written as 
\begin{eqnarray}
P_F(\bV\rightarrow \bV'| \bV,t)&=&\int \D \eta_1 \D \eta_2\,  \delta (\Delta V_1-\Delta t \cdot (f1(V_1,V_2)+\sigma_{11} \eta_1+\sigma_{12} \eta_2))\nonumber \\
&& \times \delta (\Delta V_2-\Delta t \cdot (f_2(V_1,V_2)+\sigma_{21} \eta_1+\sigma_{22} \eta_2)) p_1(\eta_1) p_2(\eta_2),\label{pf1}
\end{eqnarray} 
where $\delta(x)$ is the Dirac delta function, and $p_m(\eta_m)$ is the probability distribution of the $m$-th Gaussian noise
\begin{equation}
p_m(\eta_m)=\exp\pq{-\frac{\eta_m^2\Delta t}{4 R_m k_B T}}\sqrt{\frac{\Delta t}{4 \pi R_m k_B T_m}}.
\end{equation} 
Expressing the Dirac delta in Fourier space $\delta(x)=1/(2 \pi ) \int \D q \exp(i q x)$,
eq.~(\ref{pf1}) becomes 
\begin{eqnarray}
P_F(\bV\rightarrow \bV'| \bV,t)&=& \int \frac{\D q_1 \D q_2 }{(2  \pi)^2}\exp\pq{\ii (q_1\Delta V_1+q_2 \Delta V_2) }\int \prod_m \D \eta_m \, \E^{\Delta t \pq{\ii  q_m (f_m+\sigma_{m1}\eta_1+\sigma_{m2}\eta_2)-\frac{\eta_m^2}{4 R_m k_B T}}}\label{pf11} \\
&=& \exp\left\{-\frac{\Delta t}{4 k_B T_1 T_2 }\left[C_1^2 R_1 T_2 (\dot V_1-f_1)^2+ C_2^2 R_2 T_1 (\dot V_2 -f_2)^2 \right. \right.\nonumber \\
&&\qquad +2 C (\dot V_1 -f_1-\dot V_2 +f_2) (C_1 R_1 T_2 (\dot V_1-f_1)- C_2 R_2 T_1(\dot V_2 -f_2) ) \nonumber \\
&&\qquad \left. \left. +C^2 (R_2 T_1 +R_1 T_2) (\dot V_1 -f_1-\dot V_2 +f_2)^2\right] \right\}\frac{X}{4 \pi k_B  \Delta t} \sqrt \frac{R_1 R_2}{T_1 T_2};
\label{PFor}
\end{eqnarray} 
where we have taken $\Delta V_m/\Delta t\simeq \dot V_m$.
A similar calculation for the reverse transition gives
\begin{eqnarray}
P_R(\bV'\rightarrow \bV| \bV',t+\Delta t)&=&
\int \D \eta_1 \D \eta_2\,  \delta (\Delta V_1+\Delta t (f1(V_1',V_2')+\sigma_{11} \eta_1+\sigma_{12} \eta_2))\nonumber \\
&& \times \delta (\Delta V_2+\Delta t (f_2(V_1',V_2')+\sigma_{21} \eta_1+\sigma_{22} \eta_2)) p_1(\eta_1) p_2(\eta_2) \\
&=&   \exp\left\{-\frac{\Delta t}{4 k_B T_1 T_2 }\left[C_1^2 R_1 T_2 (\dot V_1+f_1)^2+ C_2^2 R_2 T_1 (\dot V_2 +f_2)^2 \right. \right.\nonumber \\
&&\qquad +2 C (\dot V_1 +f_1-\dot V_2 -f_2) (C_1 R_1 T_2 (\dot V_1+f_1)- C_2 R_2 T_1(\dot V_2 +f_2) ) \nonumber \\
&&\qquad \left. \left. +C^2 (R_2 T_1 +R_1 T_2) (\dot V_1 +f_1-\dot V_2 -f_2)^2\right] \right\}\frac{X}{4 \pi k_B \Delta t}\sqrt \frac{R_1 R_2}{T_1 T_2}.
\label{PBack}
\end{eqnarray} 
We now consider the ratio between the probability of the forward and backward trajectories, and by substituting the explicit definitions of $f_1(V_1,V_2)$ and $f_2(V_1,V_2)$, as given by eqs.~(\ref{eqf1})-(\ref{eqf2}), into eqs.~(\ref{PFor}) and (\ref{PBack}),  we finally obtain
\begin{equation}
\log\frac{P_F(\bV\rightarrow \bV'| \bV,t)}{P_R(\bV'\rightarrow \bV| \bV',t+\Delta t)}=-\Delta t \p{V_1 \frac{(C_1 +C) \dot V_1-C \dot V_2}{k_BT_1}+V_2\frac{(C_2 +C) \dot V_2-C \dot V_1}{k_BT_2}}=\Delta t \p{\frac{\dot Q_1}{k_BT_1}+\frac{\dot Q_2}{k_BT_2}},
\label{ratio:dt}
\end{equation} 
where we have exploited eq.~(\ref{qm}) in order to obtain the rightmost equality.
Thus, by taking a trajectory $\bV\rightarrow \bV'$ over an arbitrary time interval $[t,t+\tau]$, and by integrating the right hand side of eq.~(\ref{ratio:dt}) over such time interval, we finally obtain
\begin{eqnarray}
k_B \log\frac{P_F(\bV\rightarrow \bV'| \bV,t)}{P_R(\bV'\rightarrow \bV| \bV',t+\tau)}=\p{\frac{ Q_1}{T_1}+\frac{Q_2}{T_2}}=\Delta S_{r,\tau}
\label{sr:eq}
\end{eqnarray} 
We now note that the system is in an out-of-equilibrium steady state characterized by a PDF $P_{ss}(V_1,V_2)$, and so, along any trajectory connecting two points in the phase space $\bV$ and $\bV'$ the following equality holds 
\begin{eqnarray}
\exp\pq{\Delta S_{tot}/k_B}&=&\exp\pq{(\Delta S_{r,\tau}+\Delta S_{s,\tau)}/k_B}\nonumber \\
&=&\frac{\mathcal P_F(\bV\rightarrow \bV'| \bV,t)P_{ss}(\bV)}{ \mathcal P_R(\bV'\rightarrow \bV| \bV',t+\tau)P_{ss}(\bV')}, 
\end{eqnarray} 
where we have exploited eq.~(\ref{sr:eq}), and the definition of $\Delta S_{s,\tau}$ as given in eq.~(\ref{eq:DS_tot}).
Thus we finally obtain
\begin{eqnarray}
\mathcal P_F(\bV\rightarrow \bV'| \bV,t)P_{ss}(\bV)\exp\pq{-\Delta S_{tot}/k_B} =\mathcal P_R(\bV'\rightarrow \bV| \bV',t+\tau)P_{ss}(\bV')
\end{eqnarray} 
and summing up both sides over all the possible trajectories connecting any two points $\bV$, $\bV'$ in the phase space, and exploiting the normalization condition of the backward probability, namely
\begin{equation}
\sum_{\bV',\bV}\mathcal P_R(\bV'\rightarrow \bV| \bV',t+\tau)P_{ss}(\bV')=1, 
\label{fin:eq}
\end{equation} 
one obtains eq.~(\ref{eq:DS}).
It is worth  noting that the explicit knowledge of $P_{ss}(\bV)$ is not required in this proof.



\end{document}